\documentclass[manuscript]{aastex}
\usepackage{graphicx}
\usepackage{natbib}
\newcommand{\vlsr}{$V_{\rm LSR}$}
\newcommand{\kms}{~km~s$^{-1}$}

\slugcomment{To be submitted to ApJ}

\shorttitle{Identifying O-Rich Late/Post-AGB Stars}
\shortauthors{Yung et al.}

\begin{document}


\title{Maser and Infrared Studies of Oxygen-Rich Late/Post-AGB Stars and 
       Water Fountains: Development of a New Identification Method} 


\author{Bosco H. K. Yung\altaffilmark{1}, 
        Jun-ichi Nakashima\altaffilmark{1}$^{,}$\altaffilmark{2},
        and Christian Henkel\altaffilmark{3}$^{,}$\altaffilmark{4}
        }






\altaffiltext{1}{Department of Physics, The University of Hong Kong,
                 Pokfulam Road, Hong Kong, China}
\altaffiltext{2}{Ural Federal University, Lenin Avenue, 51, 
                 Ekaterinburg 620000, Russia}
\altaffiltext{3}{Max-Planck-Institut f{\"u}r Radioastronomie, 
                 Auf dem H{\"u}gel 69, D-53121 Bonn, Germany}
\altaffiltext{4}{Astron. Dept., King Abdulaziz University, P.O. Box 80203, 
                 Jeddah 21589, Saudi Arabia}


\begin{abstract}
We explored an efficient method to identify evolved stars with oxygen-rich 
envelopes in the late AGB or post-AGB phase of stellar evolution, which 
include a rare class of objects --- the ``water fountains''. Our method 
considers the OH and H$_{2}$O maser 
spectra, the near infrared $Q$-parameters (these are colour indices accounting 
for the effect of extinction), and far-infrared \textsl{AKARI} colours. 
Here we first present 
the results of a new survey on OH and H$_{2}$O masers. There were 108 
colour-selected objects: 53 of them were observed in the three OH 
maser lines (1612, 1665, and 1667~MHz), with 24 detections (16 new for 
1612~MHz); and 106 of them were observed in the H$_{2}$O maser line (22~GHz) 
with 24 detections (12 new). We identify a new potential water fountain 
source, IRAS~19356$+$0754, with large velocity coverages of both OH and 
H$_{2}$O maser emission. In addition, several objects with high velocity 
OH maser emission are reported for the first time. The $Q$-parameters as well 
as the infrared [09]$-$[18] and [18]$-$[65] \textsl{AKARI} 
colours of the surveyed objects are then calculated. We suggest that 
these infrared properties are effective in isolating aspherical from 
spherical objects, but the morphology may not necessarily be related to the 
evolutionary status. Nonetheless, by considering altogether the maser and 
infrared properties, the efficiency of identifying oxygen-rich 
late/post-AGB stars could be improved.
\end{abstract}





\section{Introduction}
\label{sec:intro}

The post asymptotic giant branch~(post-AGB) phase is a short transition phase 
for intermediate-mass stars (about 1 to 8~$M_{\odot}$) in the course of 
stellar evolution. It comes after the mass-losing AGB phase 
that creates thick circumstellar envelopes, and before the planetary 
nebula~(PN) phase in which those envelopes are ionized by radiation from 
exposed hot central stars and interstellar ultraviolet emission 
\citep[see,][for a review]{vanwinckel03araa}. The duration of the 
post-AGB phase could be just $\sim$30~years in case of a relatively massive 
star (i.e. close to 8~$M_{\odot}$), or up to $\sim$$5\times10^{4}$~years for a 
lower mass star \citep{blocker95aa}. Post-AGB stars are often invisible in the 
optical wavelength 
range due to their low envelope temperature ($\leq$200~K). They are usually 
observable at longer wavelengths because radiation from the stellar 
photosphere is absorbed and re-emitted by the thick envelope. As opposed to 
usual AGB stars, most of the post-AGB stars are surrounded by circumstellar
shells with aspherical morphology, showing
bipolar or even multipolar structures in high resolution infrared
images \citep[e.g.,][]{lagadec11mnras}.
\citet{sahai98aj} suggested that high velocity jets will form at the 
late AGB or early post-AGB phase which may affect the morphologies of the
remnants of the formerly ejected, approximately spherical shells. 
Studying objects with such jets is therefore an essential 
part in understanding the evolutionary process of post-AGB stars.

There are still no fully established theories on jet formation in AGB and
post-AGB stars.
One of the suggested scenarios depicts a binary system that
consists of a star with a thick envelope and its low-mass ($<$0.3~$M_{\odot}$) 
companion \citep{nordhaus06mnras,huggins07apj}. A torus is firstly formed
due to equatorially enhanced mass-loss. The jet is launched after that phase. 
The launching jet power might come from the 
magnetocentrifugal force, as is the case of young stellar objects 
\citep[e.g,][]{frank04apj}. For oxygen-rich stars
(i.e. intermediate-mass evolved stars with more oxygen than carbon 
atoms), there is a class of objects called the ``water fountains~(WFs)'' which
exhibit a special type of jets: the small in scale ($<$1000~AU), 
high velocity ($\geq 50$\kms), and strongly collimated bipolar jets. The name 
comes from the 
fact that those jets can be traced by H$_{2}$O maser emission. From the point 
of view of spectroscopy, an object is 
proposed to be a WF if its H$_{2}$O maser emission spectrum shows a larger 
velocity coverage than that of its 1612~MHz OH maser 
\citep[see][for a WF review]{imai07iaup,desmurs12iaup}. 
It is because the typical OH double-peaked line shape reveals the terminal 
expansion velocity (normally $\leq$25\kms) of the spherically expanding 
envelope formed by the AGB mass-loss. A larger spectral coverage 
of H$_{2}$O will therefore imply an outflow with velocity exceeding that of 
the spherical envelope. It is confirmed by high-angular-resolution 
interferometric observations of the 22~GHz H$_{2}$O maser line that
this phenomenon is caused by the existence of high velocity bipolar jets
\citep[e.g.,][]{imai02nature,yung11apj}.
Most of the WF H$_{2}$O maser spectra have a velocity coverage larger than
50\kms, reaching up to 200--300\kms. 
WFs are often regarded as objects which have just started to depart from 
spherical symmetry. For this reason, 
WFs are commonly seen as young post-AGB stars 
\citep[e.g,][]{suarez08apj,walsh09mnras,desmurs12iaup}. 
In addition to WFs, some OH objects (without H$_{2}$O maser detections) are 
found to have bipolar jets as well \citep{zijlstra01mnras}. These objects 
usually show rather irregular OH maser profiles with large velocity coverages 
(about 30--80\kms). The irregular profiles imply that these objects are more 
evolved than usual AGB stars \citep{deacon04apjs}. However, there are also 
exceptional cases of AGB stars with aspherical structures:
W~43A \citep{imai02nature}, X~Her \citep{hirano04apjl}, and 
V~Hya \citep{nakashima05apj} are examples of AGB stars with bipolar jets.

Even though the exact evolutionary phase for the jet to be launched is 
uncertain, it is clear from the above studies that jets play an important role 
in changing the morphology of envelopes at later stages of stellar
evolution. In order to understand the underlying process of such shaping 
mechanism, we need to obtain a larger sample of late AGB/post-AGB stars, 
especially those associated with high velocity jets such as the WFs. 
In our previous H$_{2}$O maser survey \citep[][hereafter Paper~I]{yung13apj1}, 
we have shown that 
the \textit{AKARI} two-colour diagram could be useful in selecting AGB and 
post-AGB star candidates, providing an alternative way of object selection
besides the well-known \textit{IRAS} two-colour diagram \citep{vanderveen88aa}
that has been used in many major surveys 
(see Paper~I for a summary of surveys). We have found four candidates of 
``low-velocity'' WFs, which are objects that have a small H$_{2}$O maser 
velocity coverage ($\sim$30\kms), but still fulfill the other WF criteria 
(i.e., the H$_{2}$O coverage is larger than that of OH, and the IRAS colours
are very red). 
These could be immature WFs,
but their true status has to be confirmed by observations with 
high angular resolution, e.g., very long baseline interferometric observations 
of maser lines, or infrared imagings with adaptive optics.

In this paper, we further explore potentially reliable indicators for the
evolutionary status of evolved stars, in particular for those at the 
late AGB/post-AGB phase. We have 
selected $\sim$100 evolved star candidates in the late AGB/post-AGB phase 
by using the \textit{AKARI} colour criteria 
suggested in Paper~I. Then, we searched for the 
OH and H$_{2}$O maser emission 
from these objects because those maser lines can be a probe of high velocity 
jets. Finally, we study the object distribution on the 
$Q1$--$Q2$ diagram 
\citep[which is an extinction-free infrared tool proposed by]
[see Section~\ref{ssec:q1q2}]{messineo12aa}
and also the \textit{AKARI} two-colour diagram. The former
reflects the near-infrared properties of the objects, while the latter is 
predominantly affected by the far-infrared emission. Observational details
of this work including the source selection are given in 
Section~\ref{sec:objs+obs}; the results are presented 
in Section~\ref{sec:res}, followed by the discussion in Section~\ref{sec:dis}.
Main conclusions are summarized in Section~\ref{sec:con}.

\section{Object Selection and Maser Observations}
\label{sec:objs+obs}

\subsection{Object Selection}
\label{ssec:objs}

Table~\ref{tab:objs} displays parameters of the 108 objects observed in this 
project. Most of the objects were selected from the 
\textit{AKARI} Point Source 
Catalogue \citep{kataza10akari,yamamura10akari}, according to their 
[09]$-$[18] and [18]$-$[65] colours. In Paper~I, we have defined the 
empirical ``post-AGB star'' colour region by
$0.5\leq [09]$$-$$[18]\leq 4.5$ and $-0.5\leq [18]$$-$$[65]\leq 2$, where
$[m]$$-$$[n]$=$2.5\log(F_{n}/F_{m})$,
with $F_{m}$ and $F_{n}$ representing the band fluxes at $m$ and 
$n$~$\mu$m, respectively. 
However, this region still suffers from contamination by young stellar
objects~(YSOs). Therefore, we have also checked the mid-infrared images
of all the objects during the selection, in order to exclude obvious YSO 
candidates. Evolved stars usually appear as point sources in mid-infrared 
images, e.g., in images taken with the 
\textit{Midcourse Space Experiment}~(\textit{MSX}) and 
\textit{Wide-field Infrared Survey Explorer}~(\textit{WISE}). On the contrary,
YSOs are often embedded in comparatively large scale nebulosities which show 
extended emission features at mid-infrared wavelengths. In addition to these 
new objects which were not observed before in the 1612~MHz OH nor in the 
22~GHz H$_{2}$O maser lines, some known H$_{2}$O maser sources 
(especially those newly reported in Paper~I) 
with no reported OH maser counterparts were included. 
We have also re-visited a few known WFs and low-velocity WF candidates.
The declinations of the targets are limited to $\delta > -25^{\circ}$ because
of the telescope location. Owing to limited observing time, not all objects 
were observed in both lines (see Table~\ref{tab:objs}).

\subsection{Maser Observations and Data Reduction}
\label{ssec:obs}

The OH and H$_{2}$O maser observations were carried out with the 
Effelsberg 100~m radio telescope from 2012 October 11 to 18. 
For the OH observations, an 1.3--1.7~GHz HEMT receiver was used 
with a Fast-Fourier-Transform~(FFT) spectrometer as backend. 
The frequency coverage of the spectrometer was 100~MHz. The central frequency 
was set to 1640.0~MHz so that three of the four $^2\Pi_{3/2}$ $J=3/2$, 18~cm 
$\Lambda$-doublet lines were covered. 
The adopted rest frequencies of the OH lines were 1612.2310~MHz for the 
satellite line, and 1665.4018, 1667.3590~MHz for the two main lines 
\citep{lovas04jphyschem}. The FWHM of the beam was about 
8\arcmin. With 32,768 spectral channels,  
the corresponding channel spacing for the three target 
frequencies was between 0.5\kms\ and 0.6\kms. For the H$_{2}$O observation, 
an 18--26~GHz HEMT receiver was used with another FFT spectrometer. 
The frequency coverage of the spectrometer was 500~MHz, and the central 
frequency was set to 22.235080~GHz, 
the rest frequency of the $6_{16}\rightarrow 5_{23}$ transition line of 
H$_{2}$O \citep{lovas04jphyschem}. 
The FWHM of the beam was about 40\arcsec. The number of spectral channels was
again 32,768, so the channel spacing was about 
0.8\kms. Velocity resolutions are coarser than the channel spacing, by 16\%.
The scale of the local-standard-of-rest 
velocity~(\vlsr) for both OH and H$_{2}$O observations has been confirmed to 
be accurate by comparisons with spectra from known sources. 

An ON/OFF cycle of 2~minutes was used in a position-switching mode. 
For OH masers, the OFF-position was 1$^{\circ}$ west in azimuth from the 
ON-position. For H$_{2}$O masers, the OFF-position was set to 
$-75$~s with respect to the ON-position along right ascension. This option 
was chosen because it kept the OFF-position exactly on the same track as 
the ON-position.\footnote{https://eff100mwiki.mpifr-bonn.mpg.de} 
The observing time for each source was about 6--20~minutes, with the
exception of IRAS~19356$+$0754, which was observed for 6~hours. 
The long integration time was needed to confirm the weak line components from 
this object (see Section~\ref{ssec:ind}). 
The root-mean-square~(rms) noise level ranges from
0.001 to 0.1~Jy for both OH and H$_{2}$O maser observations. Pointing was 
calibrated every 2--3 hours by doing 2-points cross scans on bright 
quasars with strong continuum emission. The typical pointing accuracy is about 
5\arcsec. 
Flux calibration was obtained using pointing sources with known 
flux densities \citep[see,][]{ott94aa}.

Data reduction were performed with the Continuum 
and Line Analysis Single-dish Software~(CLASS) 
package.\footnote{http://www.iram.fr/IRAMFR/GILDAS} 
Individual scans on each
object were inspected by eye, and those with obvious artifacts were discarded. 
The remaining scans were then averaged. The baseline of each
spectrum was fit by a low order (1st--3rd) polynomial and subtracted, using 
channels free of emission features.

\section{Results}
\label{sec:res}

\subsection{Overview of the OH and H$_{2}$O Maser Detections}
\label{ssec:oh+h2o}

Figure~\ref{fig:oh_sp} shows the spectra for objects only with OH maser 
detections, while Figure~\ref{fig:h2o_sp} is for H$_{2}$O masers only. 
For objects with both OH and H$_{2}$O masers detected, their velocity-aligned 
spectra are
shown in Figure~\ref{fig:oh+h2o_sp}. Table~\ref{tab:objs} summarizes the 
coordinates and infrared colours of the
objects that have OH and/or H$_{2}$O detections. The corresponding spectral 
parameters of all the detections and non-detections are presented in 
Tables~\ref{tab:oh} to \ref{tab:nh2o}. Amongst 108 selected objects, 53 
were observed in 
OH (which includes the 1612, 1665, and 1667~MHz maser lines), 
with 24 detections. There are 16 new 1612~MHz, 9 new 1665~MHz, and 11 new  
1667~MHz detections. Some of these newly detected lines originate from the
the same objects where previously other OH maser lines have already been
reported. For the H$_{2}$O maser line at 22~GHz, 106 objects were observed 
with 24 detections (12 new). 
The detection rates of both OH and H$_{2}$O 
masers agree with some previous maser surveys on post-AGB stars
\citep[see][for a discussion on the detection rates]{habing96aar}
, e.g., $\sim$40\% 
for OH masers \citep{hekkert96aas}, and $\sim$25\% for H$_{2}$O masers 
\citep{valdettaro01aa,deacon07apj}. 

Similar to the results of previous surveys such as those of 
\citet{hekkert89aas} and \citet{hekkert91aas}, the majority of the OH spectra 
show a double-peaked 
profile at 1612~MHz (Figure~\ref{fig:oh_sp}). This is a common characteristic 
for Type~II OH/IR stars which are classified by their IRAS colours: 
$-0.30<[12]$$-$$[25]<-0.08$ \citep{habing96aar}. The double-peaked emission 
profile features two 1612~MHz peaks that reveal the line-of-sight velocities 
of the approaching and receding sides of the spherically expanding envelope. 
The velocity halfway between the peaks is taken as the systemic velocity 
of the star. The 1665 and 1667~MHz main lines usually show similar 
double-peak profiles, but in most cases they are fainter than those of the 
1612~MHz satellite line. On the other hand,
there are three objects in our observations with only the main lines detected,
namely 1824037$+$063625, 1914408$+$114449, and IRAS~22097$+$5647.
In particular, IRAS~22097$+$5647 belongs to the Type~I OH/IR stars with 
$-0.45<[12]$$-$$[25]<-0.30$ \citep{habing96aar}. 
The OH emission profile from these stars usually shows
only the main lines; occasionally the satellite line is also found but at a 
much weaker level. The profiles of the main lines usually have an irregular 
line shape, because the main lines originate from the accelerating 
region of an expanding envelope, where the gradient in the radial velocity 
produces the irregular line shape \citep{habing96aar}.
We do not have enough information for 1824037$+$063625 and 
1914408$+$114449; they could be evolved stars or YSOs. 
Nonetheless, they are point-sources in mid-infrared images, e.g., 
the 12~$\mu$m images from the \textit{WISE} catalogue which have an angular 
resolution of $6\farcs5$ \citep{wright10aj}, suggesting that there
are no clear star forming activities around these objects.
Absorption features are found in several spectra 
(see Figure~\ref{fig:oh_sp} and Table~\ref{tab:oh}), but they may be caused by 
foreground objects \citep[e.g.,][]{hekkert96aas}, or by some emission that
contaminates the OFF-positions.

Most of the obtained H$_{2}$O maser spectra have a double-peaked  
(with some of them only showing a weak secondary feature) or an irregular 
profile (Figure~\ref{fig:h2o_sp}).
It is known that H$_{2}$O masers generally have three common emission 
profiles: single-peaked, double-peaked, and irregular 
\citep[e.g.,][]{takaba94pasj,deacon07apj}. A single-peak at the systemic 
velocity is quite commonly found in AGB stars with a lower mass-loss rate in 
the case of a spherically expanding envelope, because the masers are mainly 
tangentially amplified. When the mass-loss 
rate increases as the star evolves, the profile will likely become 
double-peaked because maser amplification along the radial direction is now 
predominant, where the two peaks come from the approaching and 
receding sides of the spherical envelope \citep{takaba94pasj}. 
However, the above justification is not 
necessarily true for all cases because maser line profiles are not solely 
governed by the mass-loss rate. An irregular profile is often seen in further 
evolved objects 
\citep[e.g., the post-AGB star IRAS~15445$-$5449,][]{sanchez11mnras} or
YSOs \citep[e.g., W51-IRS2,][]{morita92pasj} because of the development of
irregular motions possibly induced by non-spherical (e.g., bipolar) outflow 
components. Note that the H$_{2}$O 
maser emission would have a similar or slightly smaller velocity coverage than 
the OH maser in most evolved stars with the notable exception of the WFs.

\subsection{Notable Individual Objects}
\label{ssec:ind}

\subsubsection{Water Fountains}

\textsl{IRAS~18286$-$0959}. It is a known WF suggested to harbour an episodic
precessing jet which produces a ``double-helix'' jet pattern 
\citep{yung11apj}. In paper~I, we have already noticed that the H$_{2}$O maser
velocity coverage of this object has increased from 220\kms to 263\kms\ since 
its first detection \citep{deguchi07apj}. This time, more components were 
detected and the resulting coverage is now $\sim$350\kms\ 
(Figure~\ref{fig:oh+h2o_sp}). 
It is not certain
whether the jet really accelerates, or whether it is simply due to maser flux 
variation so that the new components were not detected in previous 
observations. For the OH maser, this object was reported to have a single
1612~MHz feature at 39.5\kms\ \citep{sevenster01aa}. However, here we detected 
two close peaks at $-0.2$\kms\ and $12.3$\kms. The 1612~MHz emission is 
usually stable on time scales of months or even longer. This is why the 
disappearance of the 39.5\kms\ peak and the detection
of the new components are unexpected. The reason behind is unknown, 
but it 
might hint to the fact that the OH shell has been disturbed. Possibilities
include the interference from a high velocity jet, or some turbulent motion 
due to the existence of a nearby object.

\textsl{IRAS~19134$+$2131}. The first detailed
interferometric study of the H$_{2}$O masers from this 
WF was presented by \citet{imai07apj}. Its H$_{2}$O maser spectrum used to have
three peaks at about $-120$\kms, $-40$\kms, and $-10$\kms. All of them were 
still detected in 2011 (Paper~I). This time the most blueshifted peak 
(i.e. at $-120$\kms) disappeared, and the remaining double-peak profile 
resembles those of normal AGB stars (Figure~\ref{fig:h2o_sp}). There has been 
no OH maser detection toward this object. 
A longer exposure time may be needed.

\textsl{IRAS~19356$+$0754}. This object could be a new member of the WF
class. Its OH 1612~MHz maser spectrum has many emission peaks with 
velocities ranging from $-138$\kms\ to $-71$\kms. The large velocity coverage 
($\sim$67\kms) and the irregular profile are common for very evolved objects 
like post-AGB stars or proto-planetary nebulae 
\citep[PPNe;][]{zijlstra01mnras}. The 1665 and 1667~MHz lines also have an 
irregular profile and span a similar velocity range, but the total flux is 
smaller than that of the 1612~MHz line. The H$_{2}$O maser spectrum consists 
of multiple peaks with a total velocity coverage of about 
119\kms\ (from $-145$\kms to $-26$\kms), larger than 
that of the OH masers. The H$_{2}$O lines are very weak: the
strongest peak is only about 0.2~Jy. Note that this object was observed in
H$_{2}$O before by \citet{suarez07aa}, but at that time the result was a 
non-detection
(corresponding rms $\sim 0.04$~Jy). Therefore, the maser emission could be at 
a minimum during their observing period, while another possibility is that the 
maser appeared after 2007. The systemic velocity of this object is 
about $-105$\kms\ according to the maser spectra. The kinematic distance
derived using the systemic velocity and the Galactic rotation curve 
\citep{kothes07aa} is about 30~kpc. This puts the object outside the Milky Way,
which is unlikely true. The total infrared flux estimated from its
SED is about $2.3\times 10^{-12}$~W~m$^{-2}$.
Assuming a luminosity of 10,000~$L_{\odot}$
(quite typical for a post-AGB star),
the resultant luminosity distance is about 12~kpc, but this distance also 
includes a large uncertainty.

\subsubsection{Objects with High Velocity OH Maser Emission}

\textsl{1807272$-$194639}. The OH 1612~MHz spectrum shows four emission peaks
(at about $-77$\kms, $-48$\kms, $4$\kms, and $30$\kms) 
with a maximum velocity coverage of $\sim$107\kms\ (Figure~\ref{fig:oh_sp}). 
A ``U-shaped'' double-peaked profile is found in the 1667~MHz spectrum, which 
is a signature of a spherically expanding envelope commonly associated with 
AGB stars \citep[e.g.,][]{hekkert89aas}.
The velocities of the two 1667~MHz peaks match with two of the 1612~MHz 
peaks at $-77$ and $-48$\kms. Therefore, these lines probably originate 
from the same object with systemic velocity $\sim$63\kms.
The peaks at $4$ and $30$\kms\ could be the result of a high velocity jet, 
but they could also arise from another object with a different systemic
velocity, because the OH beam covers seven more 
mid-infrared sources. The angular separations between those sources and our
target are about 100$\arcsec$ to 220$\arcsec$. In particular, the [09]$-$[18] 
colours of three of them are between 0 and 1, indicating that they could be 
AGB stars (Paper~I). Flux data for longer wavelengths are not available, and 
no other information could be found for these three 
objects. There is also an absorption feature 
found at about $-30$\kms, but it is not clear how this is related to the 
1612~MHz emission. There is no H$_{2}$O maser detection.

\textsl{IRAS~18251$-$1048}. This is a known 
OH \citep[1612 and 1667~MHz,][]{engels07aa} and H$_{2}$O \citep{engels86aa} 
maser source. It is 
characterized by a relatively wide velocity coverage ($\sim$44\kms) of its OH 
1612~MHz emission (Figure~\ref{fig:oh_sp}). The expansion velocity of the 
envelope, usually taken as half of the OH velocity coverage, is about 22\kms. 
This is amongst the largest expansion velocities for typical 
oxygen-rich AGB stars \citep{hekkert89aas}. In this observation, 
a new 1667~MHz emission peak was 
detected at about 30\kms, and the corresponding velocity coverage becomes 
$\sim$135\kms. Nonetheless, we cannot rule out the possibility of 
contamination because there is another infrared source in the vicinity 
($\sim$2\arcmin) of this object. The colour of this object is not known 
because a lot of band fluxes are missing, probably because it is very dim. 
The absorption feature at 6\kms\ looks 
suspicious as it is not usually seen in evolved stars. It might again be
explained by the presence of some foreground molecular gas, or undesired 
emissions in the OFF-positions.

\textsl{1904448$+$042318} (or \textsl{SSTGLMC~G038.3546$-$00.9519}). 
The OH 1612~MHz spectrum consists of two dominant
peaks and multiple weaker peaks (Figure~\ref{fig:oh_sp}). The total velocity
coverage is $\sim$77\kms, much larger than that of usual AGB stars. The 1665
and 1667~MHz spectra reveal a similar coverage, but instead of a dominant
double-peak, they exhibit a relatively irregular profile. 
There is another 
infrared source $1\farcm 5$ apart from this object, with colour 
[09]$-$[18]$=$0.42 (i.e., AGB candidate; note that flux data for longer 
wavelengths are not available). However, in this case we suggest that
all the OH emission is more likely arising from the same object, due to the 
fact that all the three OH lines show roughly the same velocity distribution
(from about 11 to 90\kms), indicating the same systemic velocity at about 
50\kms. It is not so likely to have two different objects producing the three
similar velocity coverages, unless they both have the same systemic velocity
and shell expansion velocity. 
If all the emission peaks originated from
the same object, then there is probably a high velocity jet which produces the
large velocity coverage. The three OH line shapes look different because in 
the case of a jet, the maser excitation would be 
mainly caused by the jet-envelope collision, which produces irregular line 
features \citep[e.g.,][]{zijlstra01mnras}. There is a suspicious absorption
feature at 20\kms\ with unknown origin. No H$_{2}$O maser is detected.

\textsl{IRAS~19027$+$0517}. There is one emission peak at about 14\kms\ and
one absorption feature at 82\kms\ in the OH 1612~MHz spectrum. The 1665 and
1667~MHz spectra have multiple peaks spreading from 14\kms\ to 86\kms\
(Figure~\ref{fig:oh_sp}). 
The resulting velocity coverage is about 72\kms, which looks like another 
candidate with high velocity outflow. 
There is no H$_{2}$O maser detection.

\subsubsection{Peculiar Detections/Non-detections}

\textsl{IRAS~18587$+$0521}. This is a new OH and 
H$_{2}$O maser source, but the spectra may originate from two objects
(Figure~\ref{fig:oh+h2o_sp}). There are two infrared sources
(IRAS~18587$+$0521A and IRAS~18587$+$0521B) with
1\arcmin\ separation under the same IRAS assignment. IRAS~18587$+$0521A falls
into the ``post-AGB star'' colour region of the aforementioned \textsl{AKARI}
two-colour diagram. There are no 65~$\mu$m data for IRAS~18587$+$0521B and 
thus its position on the diagram is unknown, but it has a bluer [09]$-$[18]
colour than its neighbour. The H$_{2}$O beam was
small enough to resolve the two sources, so the H$_{2}$O maser has been
confirmed to be arising from IRAS~18587$+$0521B. On the
contrary, there was no way to determine the origin of the OH maser. 
Therefore, even though there is an H$_{2}$O peak outside the OH 1612~MHz
coverage (i.e., WF characteristic), we do not have enough confidence to
claim it is a WF candidate. The OH spectra of this source also suffered from 
severe contamination by absorption features of unknown origin.

\textsl{IRAS~18056$-$1514}. This object was suggested to be a low-velocity WF 
after we have identified an H$_{2}$O maser peak with 0.5~Jy peak flux at 
36\kms, outside its OH maser coverage (Paper~I). That peak disappeared in 
the current observation with comparable rms noise level (30~mJy in Paper~I 
and 50~mJy here). Without that 
peak, the current spectrum looks similar to a usual AGB star 
(Figure~\ref{fig:oh+h2o_sp}). The peak disappeared probably due to 
the commonly observed flux variations of 22~GHz H$_{2}$O masers, which 
typically occur on time scales of months.

\textsl{IRAS~19312$+$1950}. The true nature of this object is still uncertain,
but it could be a post-AGB star embedded in a small dark cloud 
\citep[see,][for a detailed study of this object]{nakashima11apj}. The object
used to have two stable H$_{2}$O maser peaks at about $17$ and $48$\kms\ which 
correspond to the two tips of its bipolar flow. 
An additional peak was 
detected at $26$\kms\ in our previous survey (Paper~I).
In the present observation, we find that only the most blueshifted peak at 
$17$\kms\ is still visible (Figure~\ref{fig:oh+h2o_sp}). 
As the rms noise level of the current work is similar to that in Paper~I 
(60~mJy in Paper~I and 70~mJy here), the sudden disappearance of the other two 
emission peaks are likely due to flux variations of the H$_{2}$O maser. 
However, it might also hint at a possible change in the physical condition of 
the envelope.

\section{Discussion}
\label{sec:dis}

In this section, the $Q$-parameters and \textsl{AKARI} colours of the observed
maser sources are discussed. By adding these infrared properties, we suggest
an improved way to identify the evolutionary status of 
evolved stars, especially for those at the late AGB/post-AGB phase. 

\subsection{The $Q1$ and $Q2$ Parameters}
\label{ssec:q1q2}

The $Q1$ and $Q2$ parameters were introduced by \citet{negueruela07aa} and
\citet{messineo12aa}. They are defined as:
\begin{eqnarray}
   Q1 &=& (J-H) - 1.8 \times (H-K_{\rm s})~; \\
   Q2 &=& (J-K_{\rm s}) - 2.69 \times (K_{\rm s} - [8.0])~,\label{eq:q2}
\end{eqnarray}
where $J$~(1.25~$\mu$m), $H$~(1.65~$\mu$m), and $K_{\rm s}$~(2.17~$\mu$m)
represent the three band fluxes of the
\textit{Two Micron All Sky Survey} \citep[\textit{2MASS},][]{skrutskie06aj}, 
and [8.0] is the 8~$\mu$m band flux from the 
\textit{Galactic Legacy Infrared Mid-Plane Survey 
Extraordinaire} \citep[\textit{GLIMPSE},][]{benjamin03pasp,churchwell09pasp}.
The $Q1$ parameter was originally used to select
infrared counterparts of high-mass X-ray binaries, and it was also
useful in finding red supergiant clusters 
\citep{negueruela07aa,negueruela11aa,fok12apj}. The $Q2$ parameter was 
inspired by $Q1$, but in addition to the near-infrared \textit{2MASS} data, 
the mid-infrared [8.0] data were also included. This parameter could be used 
to measure the infrared excess due to circumstellar shells only, which is 
independent of interstellar extinction \citep{messineo12aa}. Therefore, 
comparing to $Q1$, $Q2$ is more sensitive to the nature of circumstellar
envelopes of evolved stars. In \citet{messineo12aa},
the [8.0] entries were preferably taken from the 
\textit{GLIMPSE} database, because it has a high resolution
($\sim$1\farcs 2). However, since the \textit{GLIMPSE} project mainly covered 
the region within Galactic latitudes $b=\pm 1^{\circ}$ along most part of the
Galactic plane, many of our objects are therefore not included 
because evolved stars tend to drift away from the 
Galactic plane. Thus, we have used the $A$~(8.28~$\mu$m) band data from the
\textit{Midcourse Space Experiment} \citep[\textit{MSX},][]{egan03msx} instead 
of \textit{GLIMPSE}. It was shown that this change would not 
alter the behaviour of the $Q2$ parameter 
\citep[see Figure~4 in][]{messineo12aa}. 
Since $Q1$ and $Q2$ serve like colours (but free from the effect of 
interstellar extinction) which are affected by the profile of the spectral 
energy distributions~(SEDs), it is expected that most of the objects at the 
same stage of stellar evolution will share the same ranges of $Q$ values. 
Hence, they will cluster in the $Q1$--$Q2$ diagram, similar to the cases of 
2-colour diagrams 
\citep[see, e.g., \textit{IRAS} 2-colour diagram,][]{vanderveen88aa}.
Figure~\ref{fig:q1q2_obs} shows the $Q1$--$Q2$ diagram of the observed
targets in this project, together with the H$_{2}$O sources detected in 
Paper~I. The objects are found within $-7 < Q1 <2$ and $-37 < Q2 < 0$, but 
most of them are distributed roughly in the cluster with
$-2 < Q1 <1$ and $-15 < Q2 < 0$. 
Some objects extend from this main
cluster toward the negative $Q2$ direction, while some others are found in  
another region with more negative $Q1$ values. The range of
$Q2$ values occupied by our maser sources is larger when compared to the range
of $Q1$ values, that means the objects have a wider range of flux values in 
the mid- or far- infrared regions (reflected by $Q2$), than in the 
near-infrared region (reflected by $Q1$).

Before any further discussion, we have to consider
the possible influence of artificial effects or contamination. 
The interstellar extinction has been a big problem for infrared research, and
the situation becomes even more severe for the region toward the Galactic
plane. In fact, the $Q2$ parameter was designed in a way to avoid the effect 
of extinction. Figure~\ref{fig:q2_b} shows a diagram of 
$Q2$ versus Galactic $b$. We can see that there are more OH 
maser sources in the region with $|b|<1^{\circ}$. However, 
the $Q2$ values of our OH and H$_{2}$O maser sources do not have a 
clear correlation with Galactic latitude. Hence, even if there is a 
positional dependence for the $Q2$ value, it is not significant. Another 
possible source of error comes from the resolution of the OH
maser observations. Since the large beam occasionally covered more than one
source with similar infrared characteristics, sometimes it is a bit difficult 
to confirm which source the OH maser comes from. This would not be a 
problem for the sources where H$_{2}$O masers were detected as well, because 
in those cases the origin of the OH masers could be checked by comparing the 
line velocities of both masers. The ambiguous cases are discussed in 
Section~\ref{ssec:ind} already.

Figure~\ref{fig:q1q2_all} shows basically the same $Q1$--$Q2$ diagram as
Figure~\ref{fig:q1q2_obs}, but with the addition of a number of known sources
for a comparison. The PPNe were obtained from \citet{meixner99apjs}. The PNe
were mainly selected from \citet{ruffle04mnras} and the ARVAL Catalogue of 
Bright Planetary Nebulae.\footnote{http://www.oarval.org/PNeb.htm}
The objects form three groups in
the diagram which are more or less similar to Figure~\ref{fig:q1q2_obs}. 
The group enclosed by a blue ellipse is dominated by AGB H$_{2}$O maser 
sources from Paper~I. Many of
them are still keeping a rather spherical envelope, as suggested by their 
spectral profiles. They have an elongated distribution pattern
which roughly extends along the black body curve. 
The group enclosed by a purple ellipse is dominated by 
WFs and PPNe, which are mostly bipolar objects. These objects are expected to 
be more evolved than the spherical objects, based on the assumption that jets 
usually develop at a later stage of evolution (but recall there are 
exceptions such as X~Her and V~Hya, which are AGB stars with bipolar 
structure, see Section~\ref{sec:intro}). 
Since they probably have thicker 
and more extended envelopes, the temperature is lower and hence the more
negative $Q2$ values (about $-10$ to $-40$). The position of the
new WF candidate, IRAS~19356$+$0754, is not known because we are not able to
find its $Q2$ values due to insufficient photometric data. 
The red ellipse encloses a smaller number ($\sim$20) of objects which are 
isolated from the two main groups by having 
more negative $Q1$ values (about $-7$ to $-3$). One of the WFs, OH~12.8$-$0.9, 
suggested to be a late-AGB star \citep{boboltz05apj,boboltz07apj}, 
and also the high velocity object candidate, 1904448$+$042318, are found in 
this group. The reason for their peculiar $Q1$ values is not clear.
The PNe are mainly 
found in the narrow region between the above three groups (i.e. with larger 
$Q2$ values than the PPNe), indicating a smaller mid-infrared flux than 
the PPNe. This is because the hot central star will become more exposed again 
in the PNe phase, and therefore a larger portion of the flux will be emitted 
from the star in the optical or near-infrared rather than from the 
mid-infrared. Finally, $Q2$ could not be calculated for two
representative bipolar AGB stars, X~Her and V~Hya, because there are no data 
with respect to their 8~$\mu$m flux. Their $Q1$ values are $-0.17$ and $0.08$,
respectively. From Figure~\ref{fig:q1q2_all}, we can see that they are 
unlikely to be found in the region for spherical objects, no matter what their 
$Q2$ values are. Instead, their $Q1$ values ($\sim$0) are in the middle of the 
$Q1$ range for ``bipolar objects'', which agrees with their bipolar nature.

Despite the small number of exceptional cases, we find a clear separation 
between the spherical and bipolar objects in the $Q1$--$Q2$ diagram.
Most post-AGB stars 
and even some late AGB stars are aspherical, in particular many of them show a 
certain degree of bipolarity due to jets (see Section~\ref{sec:intro}). 
The method used here by employing the $Q1$--$Q2$ diagram could isolate the 
clearly aspherical objects, which are likely 
(but not necessarily) objects at the late AGB/post-AGB phase.

\subsection{The Far-Infrared \textit{AKARI} Colours}
\label{ssec:akari}

In Paper~I, we have shown that the AGB and post-AGB stars occupy different
regions in the \textit{AKARI} [09]$-$[18] versus [18]$-$[65] two-colour 
diagram, suggesting that the \textit{AKARI} colours are useful 
for studying late stage stellar evolution. Now we can extend this work by 
considering more known objects (Figure~\ref{fig:akari}). The sample of PPNe 
and PNe is the same as that shown in Figure~\ref{fig:q1q2_all}. 

We find that the regionalization in Paper~I is over simplified because the 
objects are assumed to move along a single trend on the diagram as they
evolve. However, the object distribution in Figure~\ref{fig:akari} suggests 
the presence of two major groups of sources. Here most of the PPNe are found 
mixing with some AGB maser sources
(those assumed to have spherical envelopes) and WFs, in an elongated region 
(purple ellipse) almost parallel to the black
body curve. The AGB stars appear at the higher temperature 
end, while the WFs are at the lower temperature end. This temperature
tendency is similar to that in the $Q1$--$Q2$ diagram. However, here we have 
a stronger correlation between the PPNe and the black body curve. This is 
presumably due to the fact that the far-infrared colours are more sensitive to 
the temperature of the cold dust component, which contributes most of the flux 
from the SEDs of PPNe \citep[e.g., IRAS~16342$-$3814,][]{murakawa12aa}. 
There are some objects, which deviate from the black body curve, 
that occupy
the region with $0 <$~[09]$-$[18]~$<2$ and $0 <$~[18]$-$[65]~$<3$ 
(upper part of the green ellipse). However, there 
is no obvious sign on which kind of specific objects would behave like that. 
Nonetheless, we suspect that the aspherical objects will tend to move toward 
this region,
because the high velocity object candidate 1904448$+$042318, the known 
peculiar bipolar object IRAS~19312$+$1950 \citep{nakashima11apj}, and most of 
the selected PNe are found in this region. 

Both the envelope morphology and temperature are common indicators to
determine the evolutionary status of evolved stars. For instance, it is known
that an AGB star usually has a spherical envelope, as opposed to the 
aspherical envelope of a post-AGB star, and the envelope temperature of the 
former is higher ($\geq$300~K). However, Figure~\ref{fig:akari} seems to 
suggest that the change in temperature of the objects could be independent of 
their morphological change. That means, it is possible to have hotter 
AGB-like stars which are aspherical. In fact, as we mentioned before, W~43A, 
X~Her and V~Hya are examples of AGB stars with bipolar jets.  
If jets can be launched before the post-AGB phase, we suspect that WFs might 
not necessarily represent the short ``young post-AGB'' episode which has been
widely accepted. 
By studying the spectral energy distribution~(SED), we found that 
half of the known WFs actually have characteristics of AGB stars
(Yung et al. 2014, in prep.).
Note that on the contrary, there are almost no spherical post-AGB stars 
\citep[e.g.,][]{lagadec11mnras}. In short, the morphological change might not 
have a direct relationship to the evolutionary status: it could be safe to 
assume the cold post-AGB stars are aspherical, but it is not entirely correct 
to label all aspherical evolved stars as post-AGB stars, even though in most 
cases this is still true. Similar to the $Q1$--$Q2$ 
diagram, the \textsl{AKARI} two-colour diagram could serve the purpose of
identifying aspherical objects without assumptions on their evolutionary 
status, but the \textsl{AKARI} colours are more sensitive to the temperatures 
of the colder envelopes that mainly shine at far-infrared wavelengths.

\section{Summary and Conclusion}
\label{sec:con}

We have conducted an OH and H$_{2}$O maser survey with targets selected by
the far-infrared \textit{AKARI} colours. We found a new WF candidate, 
IRAS~19356$+$0754, and a few possible high 
velocity OH objects. New H$_{2}$O maser components were detected for the
known WF, IRAS~18286$-$0959, which might be a hint on jet acceleration.
We then studied the 
maser sources, and other known objects such as PPNe and PNe,
using the $Q1$ and $Q2$ parameters as well as the \textit{AKARI} colours.
While the $Q1$--$Q2$ diagram seems to be effective in separating the spherical 
and bipolar objects in general, the \textit{AKARI} colours show further that
the morphological change in cold sources is not necessarily related
to their evolutionary status, i.e., even though many of the aspherical objects 
are found to be post-AGB stars, some AGB stars may also develop jets before 
reaching the post-AGB phase. 
We suggest that the efficiency of identifying oxygen-rich objects during the 
late stages of stellar evolution (i.e. late AGB/post-AGB stars) 
could be improved by considering together the maser properties, the 
$Q$-parameters and the \textit{AKARI} colours.



\acknowledgments

We thank the anonymous referee for the comments and suggestions. 
This work is supported by a grant awarded to J.N. from the Research
Grants Council of Hong Kong (project code: HKU 704411P) and the Small
Project Funding of the University of Hong Kong (project code:
201007176004).
The results are based on observations with the 100~m telescope of the 
MPIfR (Max-Planck-Institut f{\"u}r Radioastronomie) at Effelsberg;
\textit{AKARI}, a JAXA project with the participation of ESA;
\textit{Two Micron All Sky Survey}, which is a joint project of the University 
of Massachusetts and the Infrared Processing and Analysis Center/California 
Institute of Technology, funded by the National Aeronautics and Space 
Administration and the National Science Foundation; and
\textit{Midcourse Space Experiment}. Processing of the data was funded by the 
Ballistic Missile Defense Organization with additional support from NASA 
Office of Space Science. This research has also made use of the NASA/IPAC 
Infrared Science Archive, which is operated by the Jet Propulsion Laboratory, 
California Institute of Technology, under contract with the National 
Aeronautics and Space Administration.






\appendix

\bibliography{ms}






\clearpage

\begin{figure}[ht]
   \figurenum{1}
   \centering
   \includegraphics[scale=0.8]{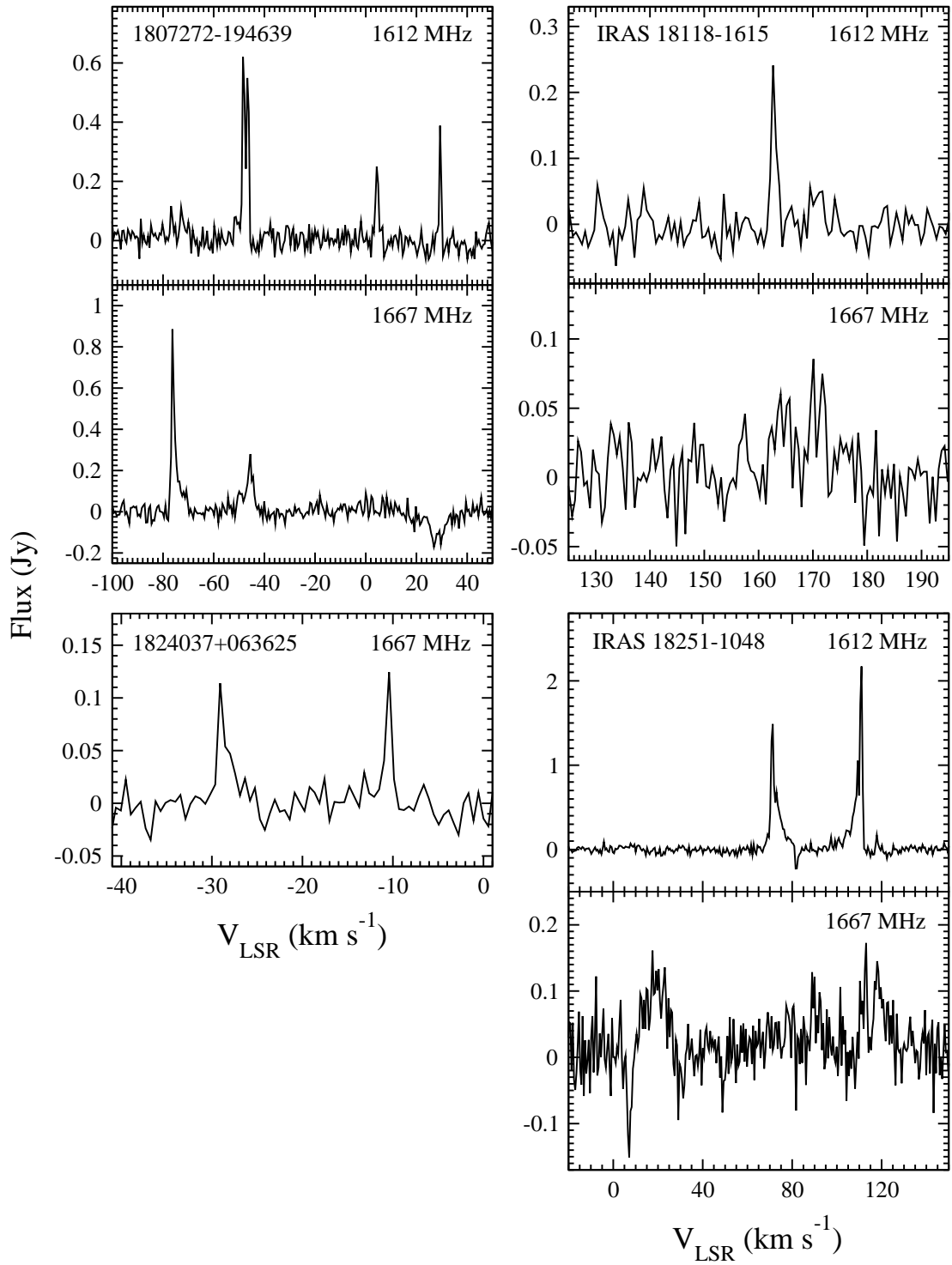}
   \caption{Spectra of objects with OH maser detections only.}
   \label{fig:oh_sp}
\end{figure}

\begin{figure}
   \figurenum{1}
   \centering
   \includegraphics[scale=0.8]{spectra_OH_02}
   \caption{\it Continued}
\end{figure}

\begin{figure}
   \figurenum{1}
   \centering
   \includegraphics[scale=0.8]{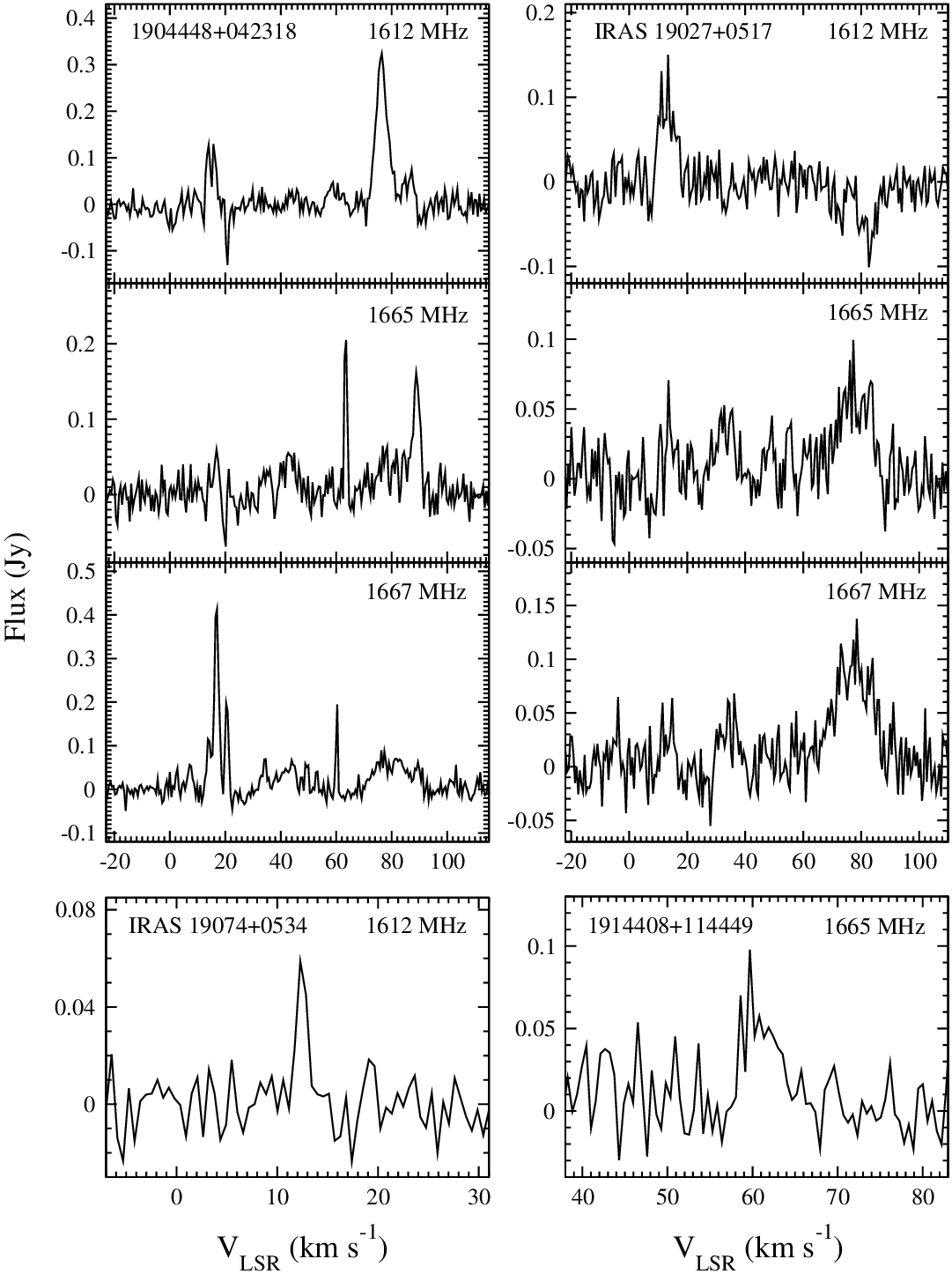}
   \caption{\it Continued}
\end{figure}

\begin{figure}
   \figurenum{1}
   \centering
   \includegraphics[scale=0.8]{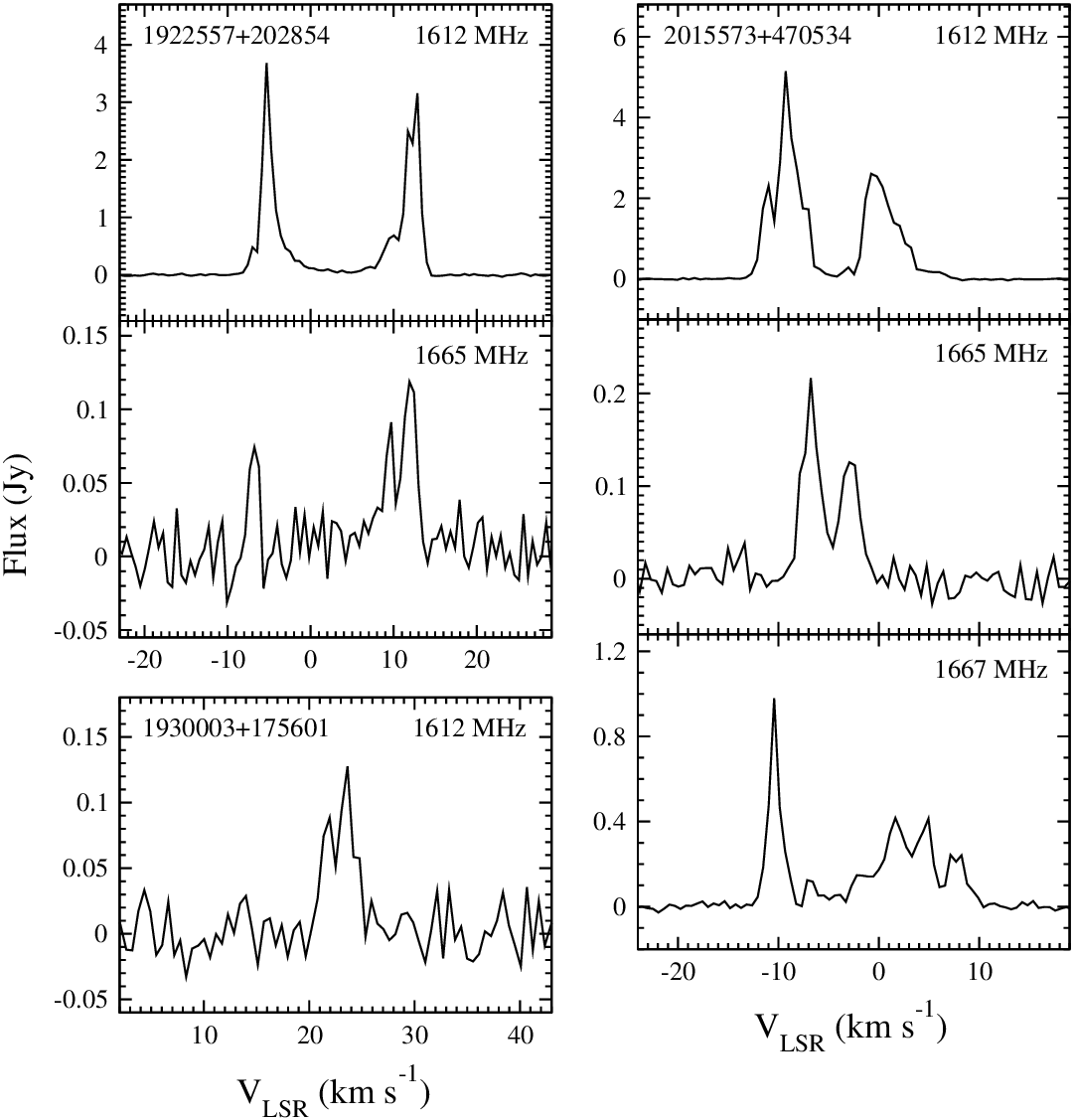}
   \caption{\it Continued}
\end{figure}


\begin{figure}[ht]
   \figurenum{2}
   \centering
   \includegraphics[scale=0.8]{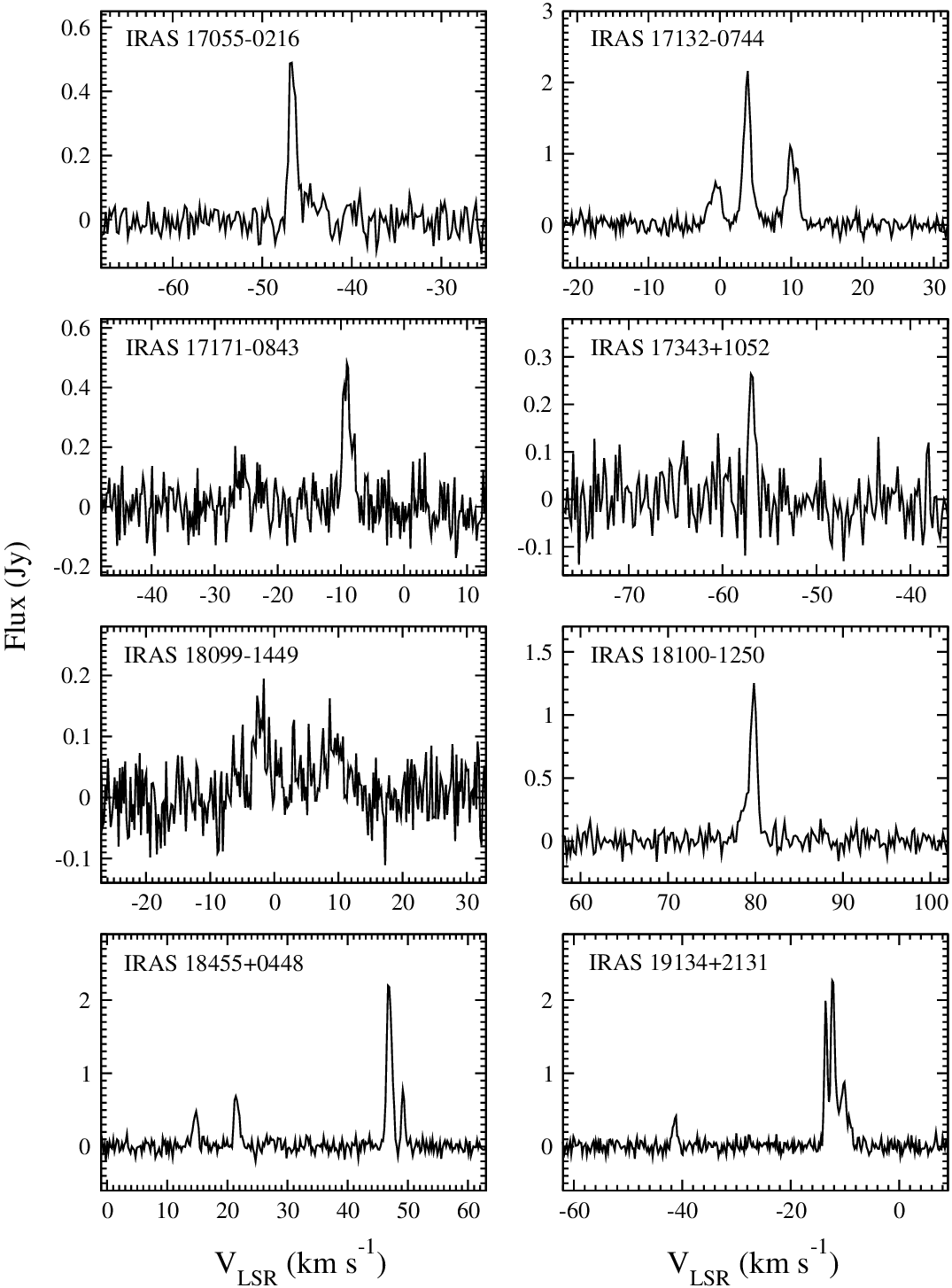}
   \caption{Spectra of objects with H$_{2}$O maser detections only.}
   \label{fig:h2o_sp}
\end{figure}

\begin{figure}[ht]
   \figurenum{2}
   \centering
   \includegraphics[scale=0.8]{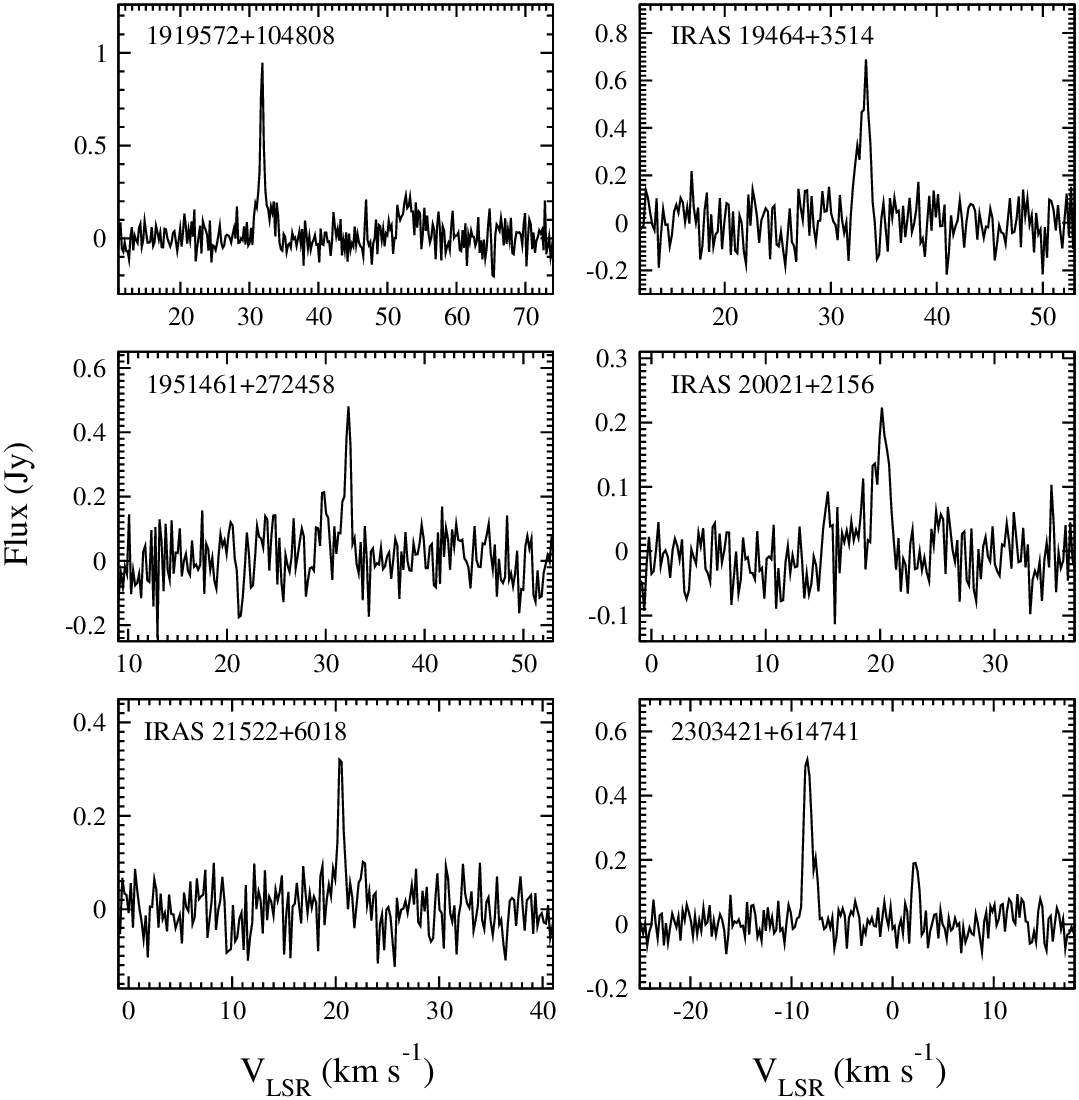}
   \caption{\it Continued}
\end{figure}


\begin{figure}[ht]
   \figurenum{3}
   \centering
   \includegraphics[scale=0.8]{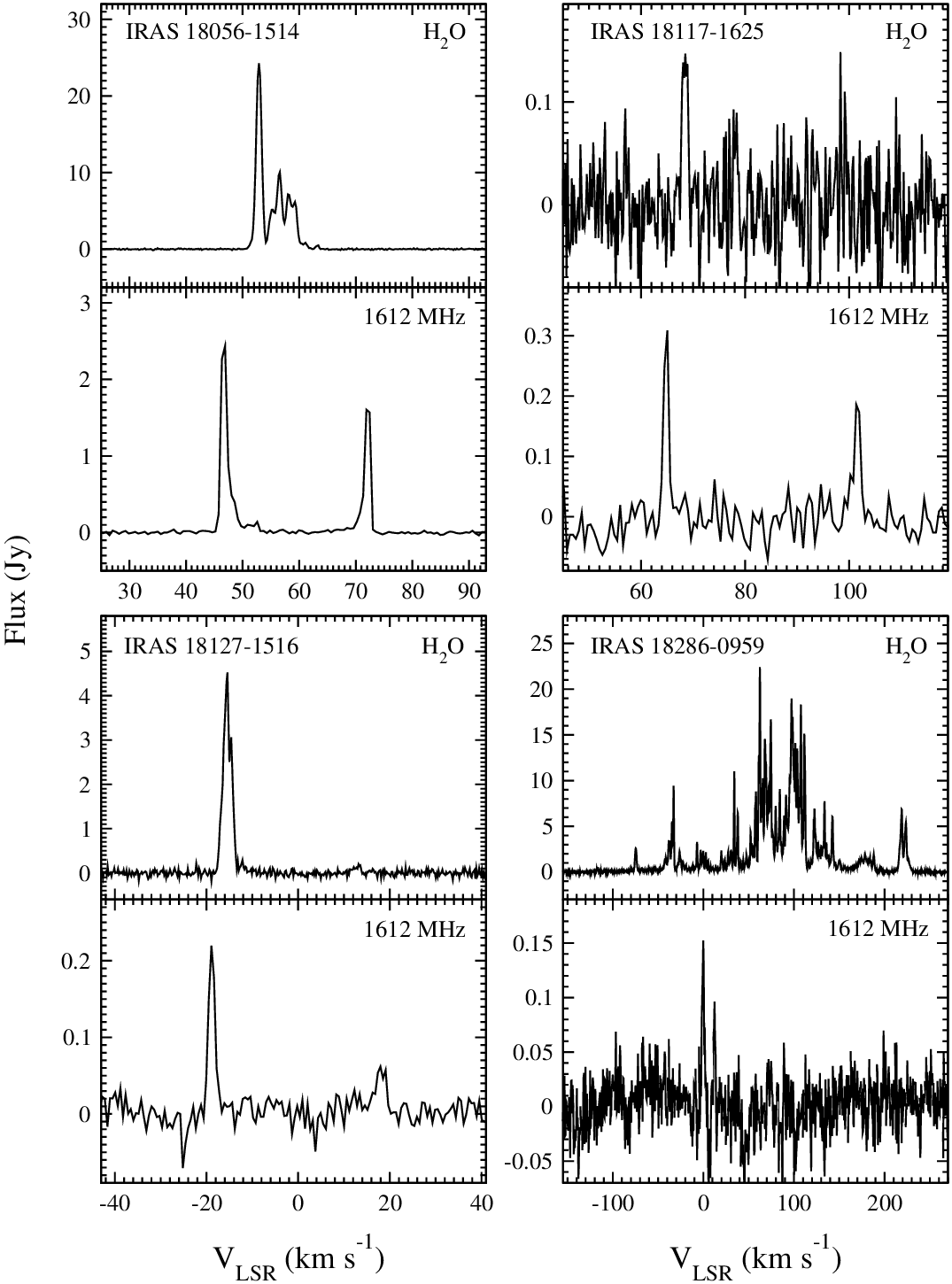}
   \caption{Spectra of objects with both OH and H$_{2}$O maser detections.}
   \label{fig:oh+h2o_sp}
\end{figure}

\begin{figure}
   \figurenum{3}
   \centering
   \includegraphics[scale=0.8]{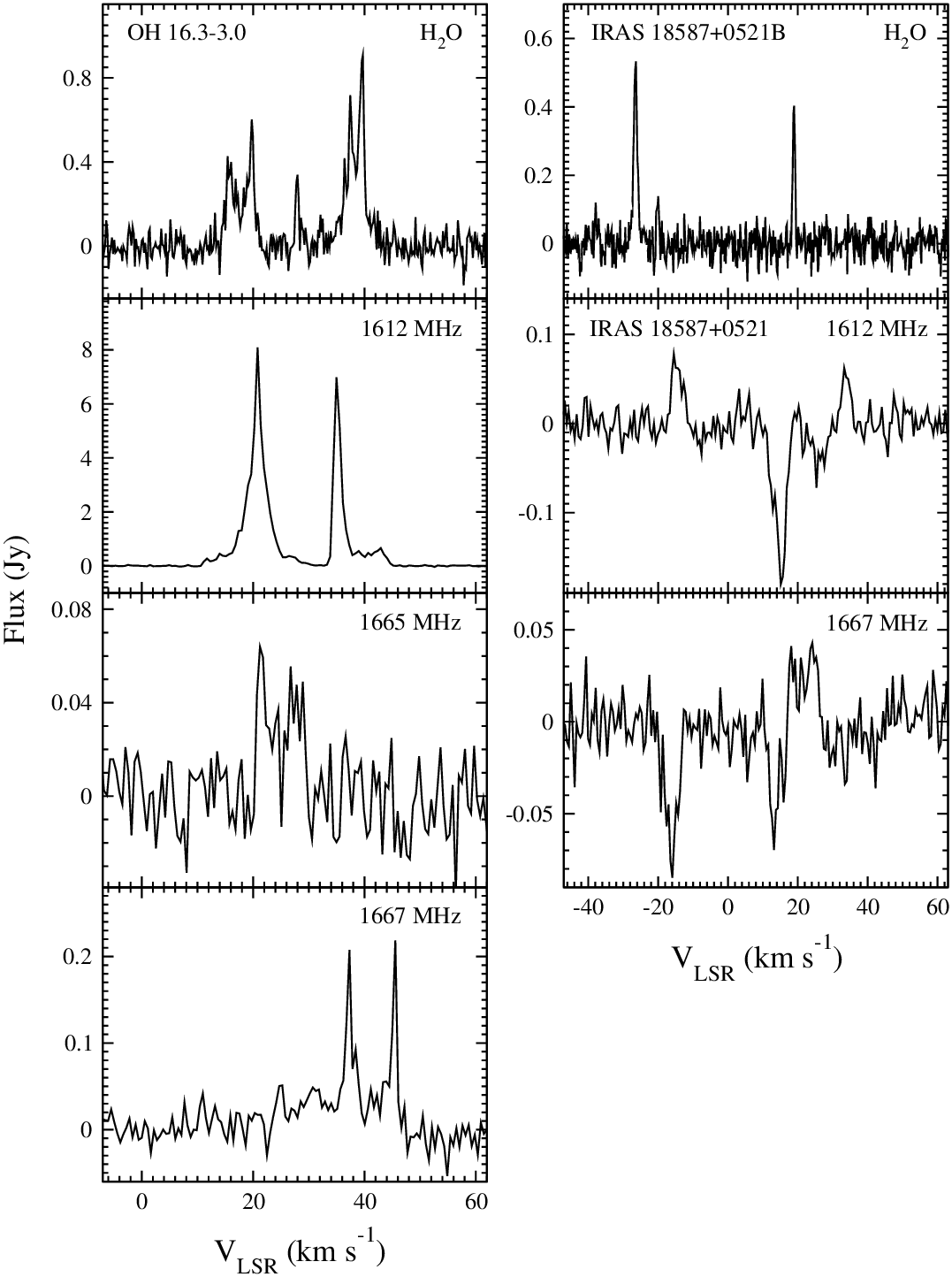}
   \caption{\it Continued}
\end{figure}

\begin{figure}
   \figurenum{3}
   \centering
   \includegraphics[scale=0.8]{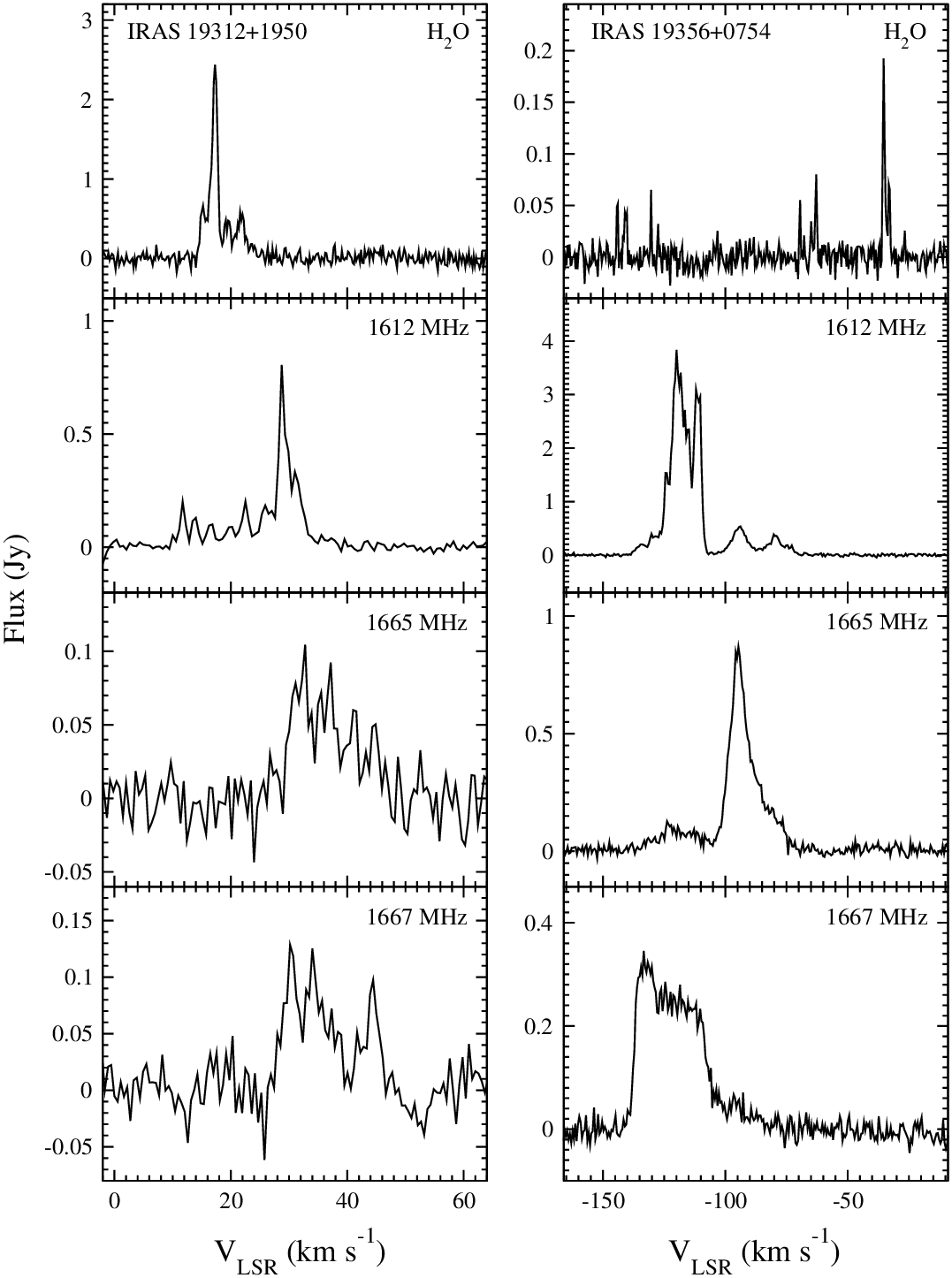}
   \caption{\it Continued}
\end{figure}

\begin{figure}
   \figurenum{3}
   \centering
   \includegraphics[scale=0.8]{spectra_H2O+OH_04}
   \caption{\it Continued}
\end{figure}



\begin{figure}[ht]
   \figurenum{4}
   \centering
   \includegraphics[scale=0.9]{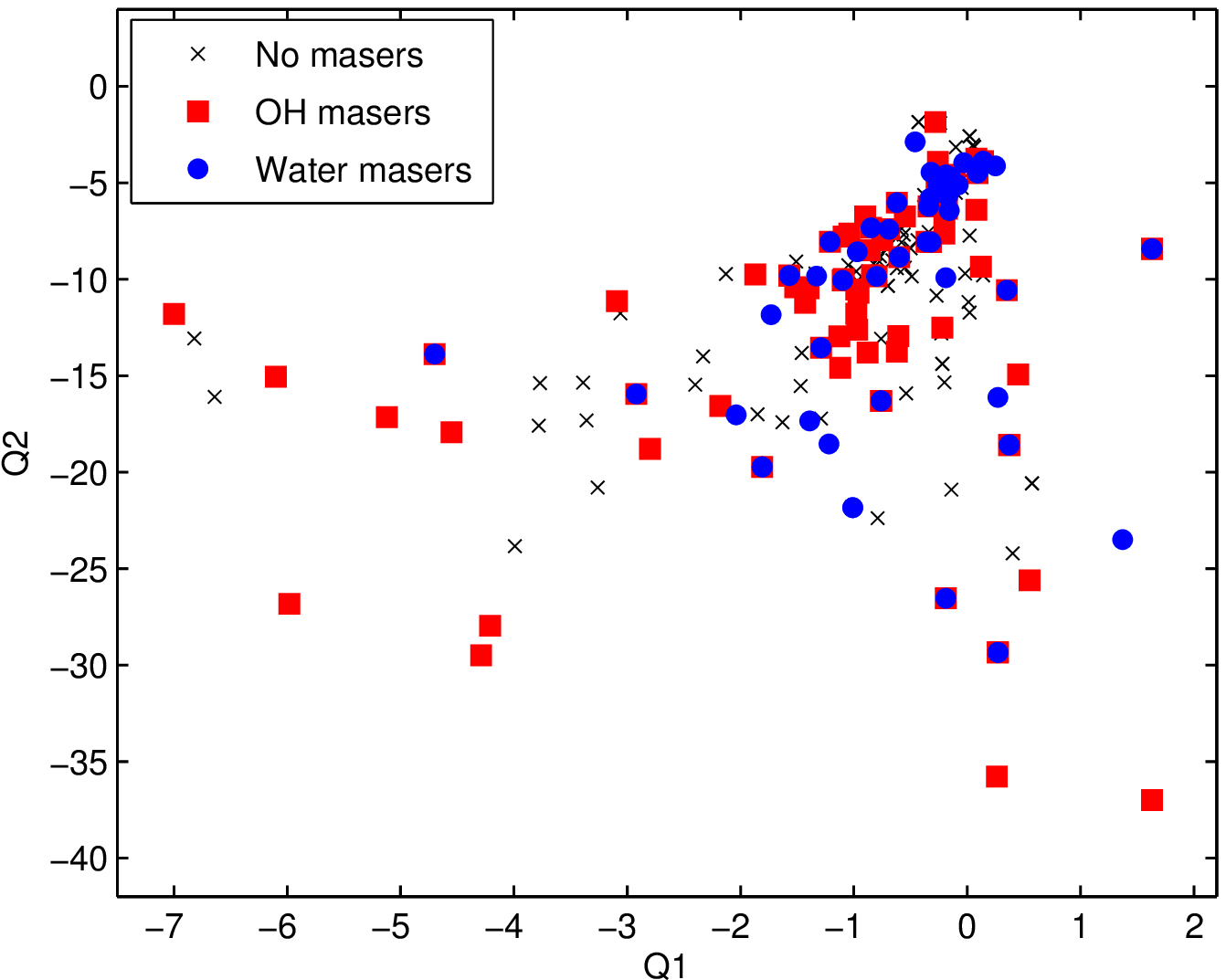}
   \caption{$Q1$--$Q2$ diagram of the objects observed in the current 
            project. The OH and H$_{2}$O sources are specified by the
            filled red squares and filled blue circles, respectively. 
            The H$_{2}$O detections in Paper~I are also included.  
            Many of them are AGB stars. The non-detections are
            specified by the black crosses.
           }
   \label{fig:q1q2_obs}
\end{figure}



\begin{figure}[ht]
   \figurenum{5}
   \centering
   \includegraphics[scale=0.9]{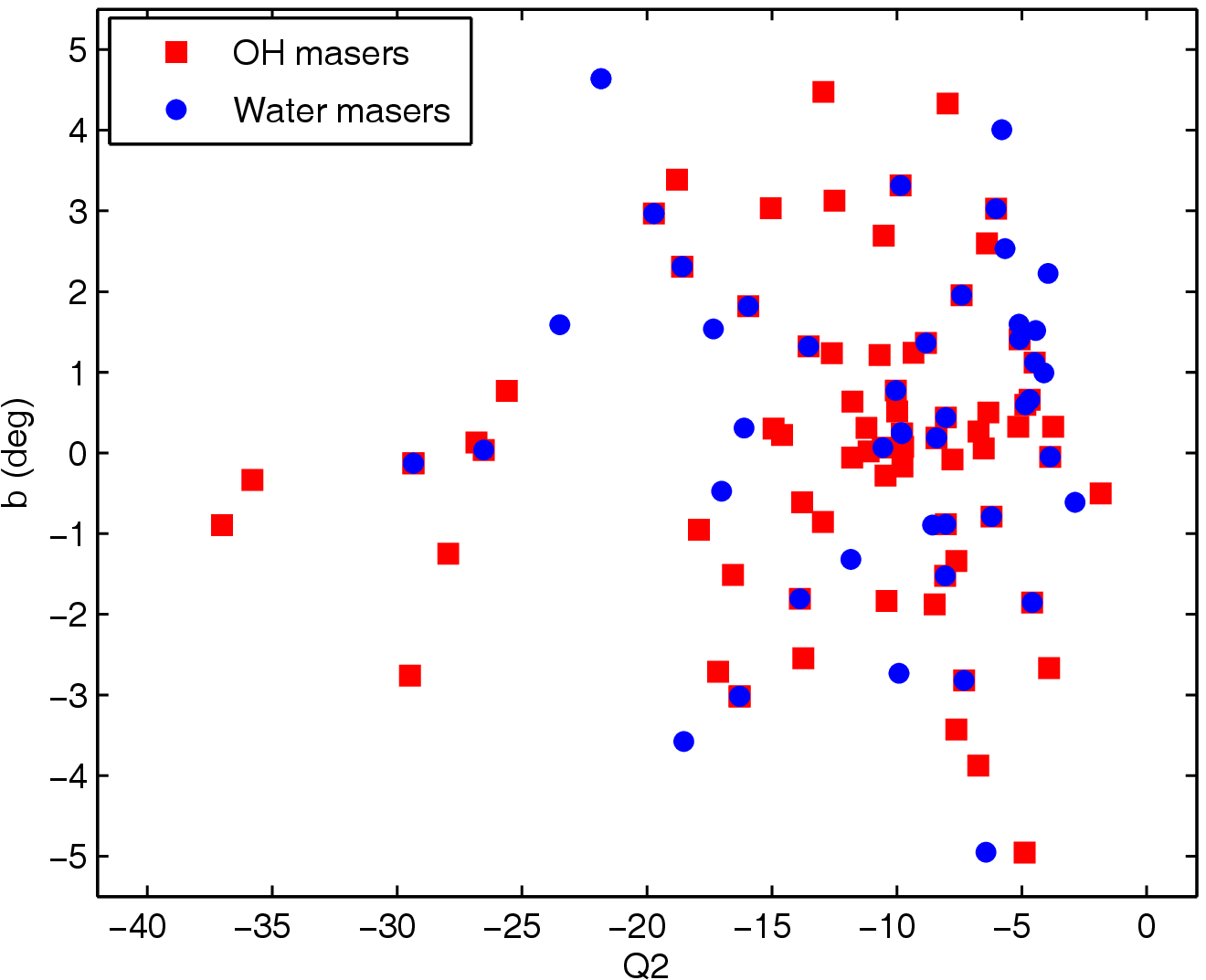}
   \caption{Galactic latitude, $b$, versus $Q2$ (see eq.~(\ref{eq:q2})) of the 
            OH and H$_{2}$O sources detected in the current observation 
            or Paper~I.
           }
   \label{fig:q2_b}
\end{figure}



\begin{figure}[ht]
   \figurenum{6}
   \centering
   \includegraphics[scale=0.9]{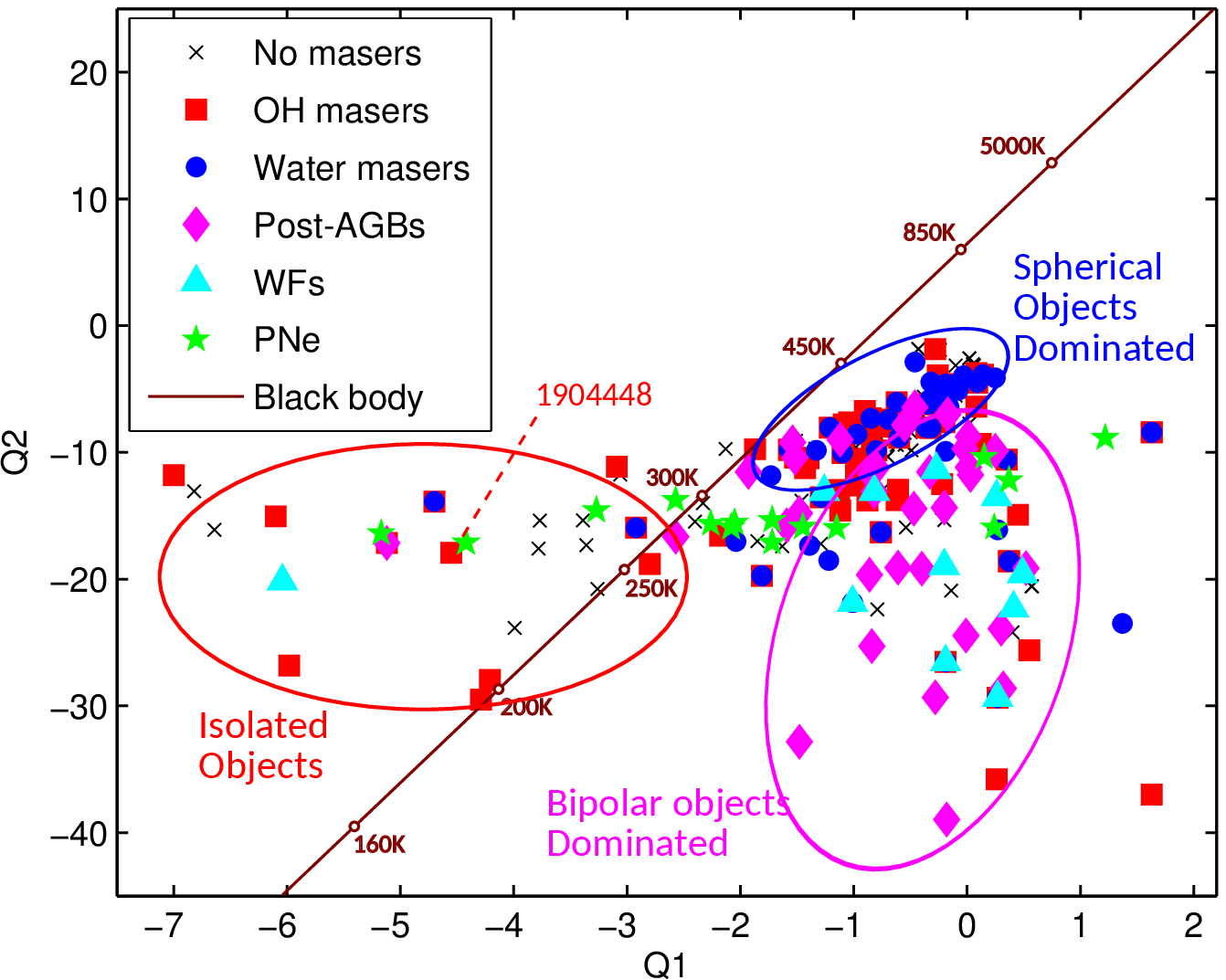}
   \caption{Same as Figure~\ref{fig:q1q2_obs}, but more sources such as WFs,
            PPNe, and PNe, are included. The objects fall into three regions.
            One is dominated by objects with spherical envelopes (enclosed by
            the blue ellipse);
            another one is dominated by objects associated with bipolar jets
            (enclosed by the purple ellipse); the third one contains 
            the ``isolated objects'' (enclosed by the red ellipse).
            The reason for their peculiar Q values is not clear 
            (see Section~\ref{ssec:q1q2}).
            The black body curve is shown in brown.
           }
   \label{fig:q1q2_all}
\end{figure}



\begin{figure}[ht]
   \figurenum{7}
   \centering
   \includegraphics[scale=0.9]{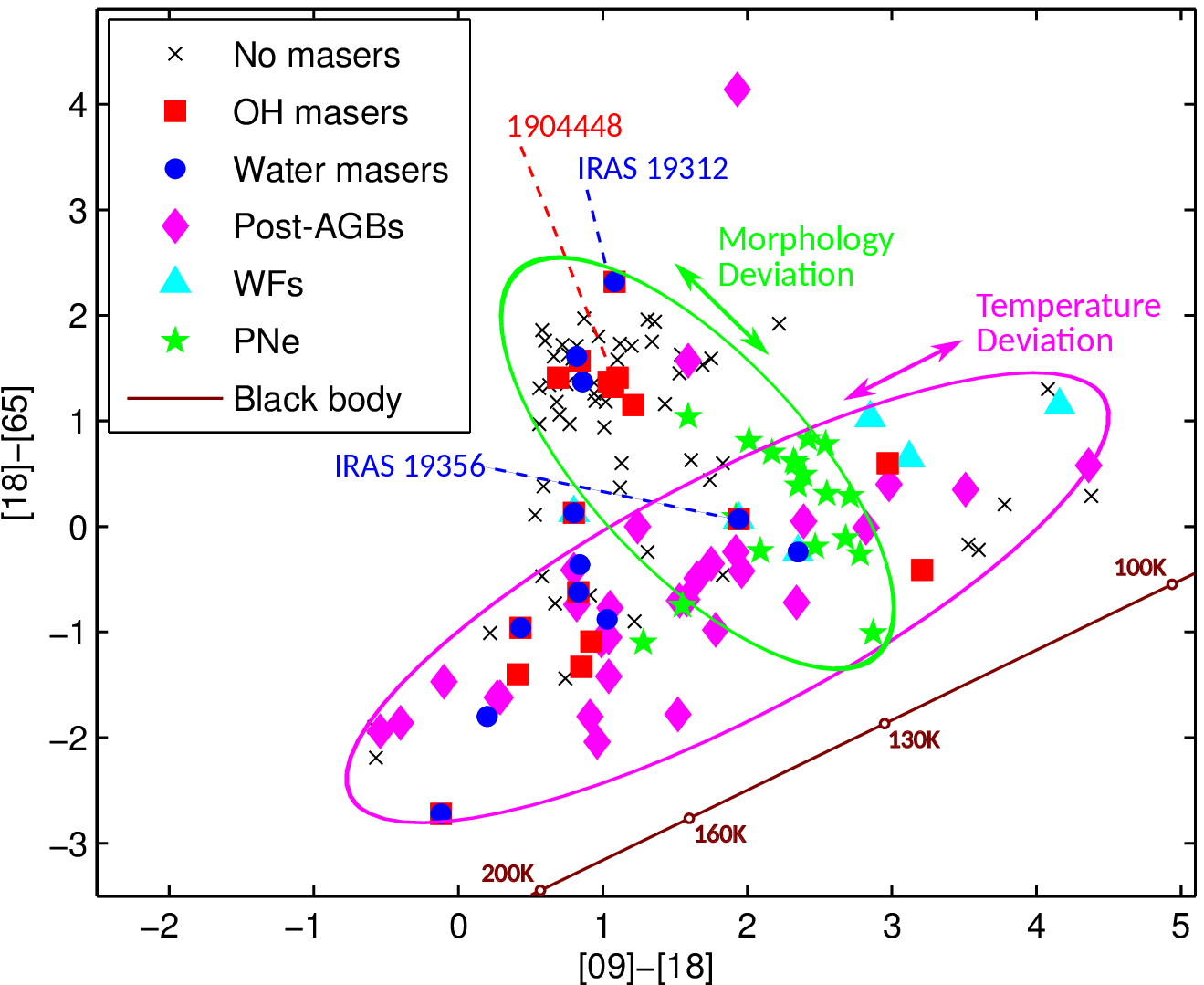}
   \caption{AKARI two-color diagram of the targets observed in
            the current project, the H$_{2}$O sources in Paper~I, and some
            other known sources (same sample as in Figure~\ref{fig:q1q2_all}). 
            The objects fall into two elongated regions. 
            The region enclosed by the purple ellipse shows clear correlation 
            with the black body temperature (shown in brown), while the one
            enclosed by the green ellipse seems to be related
            to the morphologies (e.g. spherical or bipolar). 
           }
   \label{fig:akari}
\end{figure}








\clearpage

\begin{deluxetable}{llrrrrrrrr}
\tablewidth{0pt}
\tabletypesize{\scriptsize}
\tablecaption
{Parameters of the objects observed.\label{tab:objs}}
\tablehead{


   \colhead{Object} &
   \colhead{Other Name} &
   \colhead{R.A.\tablenotemark{a}} &
   \colhead{Decl.\tablenotemark{a}} &
   \colhead{AC$_{12}$\tablenotemark{b}} &
   \colhead{AC$_{23}$\tablenotemark{b}} &
   \colhead{$Q1$\tablenotemark{c}} &
   \colhead{$Q2$\tablenotemark{c}} &
   \colhead{OH\tablenotemark{d}} &
   \colhead{H$_{2}$O\tablenotemark{d}} 

}
\startdata
0038592$+$592746 & INS2001 J003859.4$+$592749 & 00 38 59.28 & $+$59 27 46.9 & 0.75 & 1.35 & $-$0.49 & $-$9.84 & $\cdots$ & N \\
0122182$+$634055 & $\cdots$ & 01 22 18.28 & $+$63 40 55.3 & 0.56 & 1.31 & $-$0.55 & $-$9.39 & $\cdots$ & N \\
0128404$+$632737 & $\cdots$ & 01 28 40.47 & $+$63 27 37.9 & 0.60 & 1.76 & $-$0.62 & $-$9.42 & $\cdots$ & N \\
0253373$+$691539 & IRAS~02490$+$6903 & 02 53 37.39 & $+$69 15 39.2 & 1.54 & 1.63 & $-$4.10 & $\cdots$ & $\cdots$ & N \\
0358076$+$624425 & $\cdots$ & 03 58 07.65 & $+$62 44 25.8 & 0.70 & 1.06 & $-$1.13 & $\cdots$ & $\cdots$ & N \\
0413220$+$501428 & IRAS~04096$+$5006 & 04 13 22.07 & $+$50 14 28.2 & 0.66 & 1.61 & $-$1.34 & $-$9.70 & $\cdots$ & N \\
0433014$+$343840 & IRAS~04297$+$3432 & 04 33 01.45 & $+$34 38 40.6 & 1.12 & 0.37 & $-$0.78 & $\cdots$ & $\cdots$ & N \\
0536468$+$314600 & IRAS~05335$+$3144 & 05 36 46.88 & $+$31 46 00.2 & 0.94 & 1.36 & 0.02 & $-$7.73 & $\cdots$ & N \\
0540502$+$340241 & $\cdots$ & 05 40 50.22 & $+$34 02 41.3 & 0.82 & 1.71 & 0.16 & $\cdots$ & $\cdots$ & N \\
0547450$+$003842 & 2MASS~J05474500$+$0038418 & 05 47 45.00 & $+$00 38 42.1 & 1.02 & 1.18 & $-$0.50 & $\cdots$ & $\cdots$ & N \\
0608452$+$130841 & $\cdots$ & 06 08 45.28 & $+$13 08 41.3 & 1.53 & 1.45 & $-$3.06 & $-$11.76 & $\cdots$ & N \\
IRAS~16030$-$0634 & $\cdots$ & 16 05 46.33 & $-$06 42 27.9 & $-$0.46 & $\cdots$ & $-$0.16 & $\cdots$ & N & N \\
IRAS~16131$-$0216 & $\cdots$ & 16 15 47.66 & $-$02 23 31.9 & $-$0.64 & $\cdots$ & 0.28 & $\cdots$ & N & N \\
IRAS~17055$-$0216 & $\cdots$ & 17 08 10.20 & $-$02 20 21.0 & $-$0.22 & $\cdots$ & $-$0.63 & $\cdots$ & N & Y \\
IRAS~17132$-$0744 & $\cdots$ & 17 15 56.40 & $-$07 47 33.0 & $\cdots$ & $\cdots$ & $\cdots$ & $\cdots$ & N & Y \\
IRAS~17171$-$0843 & $\cdots$ & 17 19 53.45 & $-$08 46 59.7 & $-$0.11 & $\cdots$ & $-$0.70 & $\cdots$ & $\cdots$ & Y \\
IRAS~17193$-$0601 & $\cdots$ & 17 22 02.30 & $-$06 04 13.0 & $-$0.46 & $\cdots$ & $-$0.05 & $\cdots$ & N & N \\
IRAS~17343$+$1052 & $\cdots$ & 17 36 44.45 & $+$10 51 07.0 & $-$0.72 & $\cdots$ & 0.01 & $\cdots$ & $\cdots$ & Y \\
1741385$-$241435 & IRAS~17385$-$2413 & 17 41 38.52 & $-$24 14 35.8 & 1.22 & $-$0.90 & 0.40 & $-$24.21 & $\cdots$ & N \\
1750356$-$203743 & IRAS~17476$-$2036 & 17 50 35.60 & $-$20 37 43.5 & 0.58 & $-$0.47 & $-$0.72 & $-$7.88 & N & N \\
1752536$-$184100 & $\cdots$ & 17 52 53.68 & $-$18 41 00.8 & 0.74 & $-$1.44 & $-$6.64 & $-$16.10 & $\cdots$ & N \\
1807272$-$194639 & IRAS~18044$-$1947 & 18 07 27.26 & $-$19 46 39.3 & 0.92 & $-$1.09 & $-$1.43 & $-$11.22 & Y & N \\
IRAS~18056$-$1514 & $\cdots$ & 18 08 28.40 & $-$15 13 30.0 & 0.43 & $-$0.96 & 0.37 & $-$18.59 & Y & Y \\
IRAS~18099$-$1449 & $\cdots$ & 18 12 47.37 & $-$14 48 50.0 & $-$0.68 & $\cdots$ & $-$0.08 & $-$5.11 & N & Y \\
IRAS~18100$-$1250 & $\cdots$ & 18 12 50.49 & $-$12 49 44.8 & $-$0.39 & $\cdots$ & $-$0.17 & $-$5.68 & N & Y \\
IRAS~18117$-$1625 & $\cdots$ & 18 14 38.70 & $-$16 24 39.0 & 0.04 & $\cdots$ & $-$1.21 & $-$8.05 & Y & Y \\
IRAS~18118$-$1615 & $\cdots$ & 18 14 41.35 & $-$16 14 03.0 & 0.47 & $\cdots$ & $-$1.09 & $-$9.99 & Y & N \\
IRAS~18127$-$1516 & $\cdots$ & 18 15 39.90 & $-$15 15 13.0 & $\cdots$ & $\cdots$ & $-$1.10 & $-$10.05 & Y & Y \\
1824037$+$063625 & IRAS~18216$+$0634 & 18 24 03.72 & $+$06 36 25.8 & 0.85 & $-$1.33 & $-$1.19 & $\cdots$ & Y & N \\
1824288$-$155108 & IRAS~18216$-$1552 & 18 24 28.88 & $-$15 51 09.0 & 0.22 & $-$1.01 & $-$3.39 & $-$15.35 & $\cdots$ & N \\
IRAS~18251$-$1048 & $\cdots$ & 18 27 56.30 & $-$10 46 58.0 & 0.80 & $\cdots$ & $-$1.57 & $-$9.80 & Y & $\cdots$ \\
IRAS~18286$-$0959 & $\cdots$ & 18 31 22.93 & $-$09 57 21.7 & 0.80 & 0.13 & 0.27 & $-$29.35 & Y & Y \\
OH~16.3$-$3.0     & TVH89~313 & 18 31 31.51 & $-$16 08 46.5 & 0.83 & $-$0.62 & $-$0.76 & $-$16.30 & Y & Y \\
IRAS~18362$-$0521 & $\cdots$ & 18 38 57.47 & $-$05 18 28.0 & $\cdots$ & $\cdots$ & $-$0.10 & $-$5.51 & $\cdots$ & N \\
1838595$-$052024 & IRAS~18363$-$0523 & 18 38 59.57 & $-$05 20 24.0 & 0.41 & $-$1.40 & $-$0.27 & $-$5.15 & Y & N \\
1848010$+$000448 & IRAS~18454$+$0001 & 18 48 01.09 & $+$00 04 48.6 & 4.08 & 1.30 & $-$0.19 & $-$12.53 & $\cdots$ & N \\
IRAS~18455$+$0448 & $\cdots$ & 18 48 02.30 & $+$04 51 30.5 & 1.03 & $-$0.88 & $-$1.81 & $-$19.73 & $\cdots$ & Y \\
1854158$+$011501 & $\cdots$ & 18 54 15.85 & $+$01 15 01.8 & 0.84 & 1.57 & $-$7.00 & $-$11.79 & Y & N \\
1854250$+$004958 & MSX6C~G034.0126$-$00.2832 & 18 54 25.10 & $+$00 49 58.2 & 0.69 & 1.41 & $-$1.40 & $-$10.46 & Y & N \\
IRAS~18587$+$0521A & $\cdots$ & 19 01 08.43 & $+$05 25 48.0 & $-$0.05 & $\cdots$ & 0.08 & $-$3.75 & $\cdots$ & N \\
IRAS~18587$+$0521 & $\cdots$ & 19 01 10.70 & $+$05 25 46.0 & $\cdots$ & $\cdots$ & 0.58 & $\cdots$ & Y & $\cdots$ \\
IRAS~18587$+$0521B & $\cdots$ & 19 01 12.40 & $+$05 25 43.4 & 1.95 & $\cdots$ & 0.27 & $-$16.12 & $\cdots$ & Y \\
IRAS~18596$+$0605 & $\cdots$ & 19 02 04.69 & $+$06 10 09.5 & 0.04 & $\cdots$ & $-$0.28 & $-$5.78 & $\cdots$ & N \\
1904448$+$042318 & SSTGLMC~G038.3546$-$00.9519 & 19 04 44.90 & $+$04 23 18.2 & 1.10 & 1.41 & $-$4.55 & $-$17.93 & Y & N \\
IRAS~19027$+$0517 & $\cdots$ & 19 05 14.28 & $+$05 21 52.2 & 0.35 & $\cdots$ & $-$0.88 & $-$13.79 & Y & N \\
IRAS~19074$+$0534 & $\cdots$ & 19 09 54.81 & $+$05 39 06.9 & 1.04 & 1.37 & $-$2.18 & $-$16.56 & Y & N \\
IRAS~19085$+$1038 & $\cdots$ & 19 10 57.20 & $+$10 43 38.0 & $\cdots$ & $\cdots$ & 0.19 & $\cdots$ & N & N \\
1911358$+$133111 & IRAS~19092$+$1326 & 19 11 35.85 & $+$13 31 11.1 & 1.74 & 0.44 & $-$1.85 & $-$17.00 & N & N \\
1912477$+$033435 & $\cdots$ & 19 12 47.75 & $+$03 34 35.8 & 3.78 & 0.21 & $-$0.23 & $-$12.80 & N & N \\
1914408$+$114449 & IRAS~19123$+$1139 & 19 14 40.83 & $+$11 44 49.4 & 1.21 & 1.15 & $-$0.90 & $-$6.73 & Y & N \\
IRAS~19134$+$2131 & $\cdots$ & 19 15 35.19 & $+$21 36 33.6 & 2.35 & $-$0.24 & $-$1.01 & $-$21.84 & N & Y \\
1918205$+$014659 & V\*~V605~Aql & 19 18 20.57 & $+$01 46 59.0 & 1.83 & 0.60 & $-$0.43 & $\cdots$ & N & N \\
1919572$+$104808 & IRAS~19175$+$1042 & 19 19 57.24 & $+$10 48 08.8 & 0.84 & $-$0.36 & $-$1.73 & $-$11.84 & $\cdots$ & Y \\
1922250$+$131851 & IRAS~19201$+$1313 & 19 22 25.08 & $+$13 18 51.6 & 0.87 & 1.97 & $-$2.33 & $-$14.00 & $\cdots$ & N \\
1922557$+$202854 & IRAS~19207$+$2023 & 19 22 55.78 & $+$20 28 54.3 & 3.21 & $-$0.41 & 0.08 & $-$6.40 & Y & N \\
1923002$+$151051 & IRAS~19207$+$1504 & 19 23 00.28 & $+$15 10 51.3 & 0.67 & $-$0.73 & $-$1.47 & $-$15.53 & $\cdots$ & N \\
1930003$+$175601 & MSX6C~G053.2176$-$00.0808 & 19 30 00.30 & $+$17 56 01.7 & 1.07 & 1.32 & $-$1.09 & $-$7.78 & Y & N \\
IRAS~19312$+$1950 & $\cdots$ & 19 33 24.30 & $+$19 56 55.0 & 1.08 & 2.32 & 1.63 & $-$8.42 & Y & Y \\
IRAS~19356$+$0754 & $\cdots$ & 19 38 01.19 & $+$08 01 33.0 & 1.94 & 0.07 & $-$1.13 & $\cdots$ & Y & Y \\
1938574$+$103016 & IRAS19365+1023 & 19 38 57.42 & $+$10 30 16.0 & 0.53 & 0.11 & $-$0.23 & $\cdots$ & N & N \\
IRAS~19464$+$3514 & $\cdots$ & 19 48 15.96 & $+$35 22 06.1 & 0.20 & $-$1.80 & $-$1.17 & $\cdots$ & $\cdots$ & Y \\
1951461$+$272458 & 2MASS~J19514615$+$2724587 & 19 51 46.20 & $+$27 24 58.5 & 0.82 & 1.61 & $-$1.33 & $-$9.83 & N & Y \\
IRAS~20010$+$2508 & $\cdots$ & 20 03 08.30 & $+$25 17 27.0 & $-$0.75 & $\cdots$ & $-$0.24 & $-$1.90 & N & N \\
2003357$+$284847 & IRAS~20015$+$2840 & 20 03 35.71 & $+$28 48 47.2 & 1.75 & 1.59 & $-$0.65 & $\cdots$ & N & N \\
2003599$+$351617 & $\cdots$ & 20 03 59.97 & $+$35 16 17.9 & 0.58 & 1.86 & $-$0.50 & $-$8.39 & N & N \\
IRAS~20021$+$2156 & $\cdots$ & 20 04 17.30 & $+$22 04 59.0 & $-$0.27 & $\cdots$ & $-$0.16 & $-$6.44 & N & Y \\
2008383$+$410040 & IRAS~20068$+$4051 & 20 08 38.39 & $+$41 00 40.4 & 2.22 & 1.92 & 0.14 & $-$9.79 & N & N \\
2009217$+$271859 & IRAS~20072$+$2710 & 20 09 21.72 & $+$27 18 59.2 & 0.94 & 1.26 & $-$0.56 & $-$7.68 & $\cdots$ & N \\
2013142$+$370536 & IRAS~20113$+$3656 & 20 13 14.22 & $+$37 05 36.9 & 1.01 & 0.94 & $-$0.77 & $-$8.80 & $\cdots$ & N \\
2015573$+$470534 & IRAS~20144$+$4656 & 20 15 57.33 & $+$47 05 34.5 & 2.97 & 0.60 & $-$1.11 & $\cdots$ & Y & N \\
2020151$+$364334 & $\cdots$ & 20 20 15.14 & $+$36 43 34.5 & 1.10 & 1.58 & $-$0.98 & $-$9.59 & $\cdots$ & N \\
IRAS~20215$+$6243 & $\cdots$ & 20 22 20.05 & $+$62 53 02.2 & $-$0.56 & $\cdots$ & $-$0.08 & $\cdots$ & N & N \\
2029222$+$403543 & IRAS~20275$+$4025 & 20 29 22.22 & $+$40 35 43.5 & 1.20 & 1.71 & $-$2.40 & $-$15.46 & $\cdots$ & N \\
IRAS~20305$+$6246 & $\cdots$ & 20 31 26.54 & $+$62 56 49.8 & $-$0.89 & $\cdots$ & $-$0.22 & $\cdots$ & N & N \\
2033464$+$450840 & MSX6C~G083.3609$+$02.9902 & 20 33 46.48 & $+$45 08 40.4 & 1.31 & $-$0.24 & $-$1.63 & $-$17.41 & $\cdots$ & N \\
2040444$+$465322 & IRAS~20390$+$4642 & 20 40 44.45 & $+$46 53 22.4 & 0.56 & 0.97 & $-$0.97 & $-$10.08 & N & N \\
2048044$+$390459 & IRAS~20461$+$3853 & 20 48 04.41 & $+$39 04 59.7 & 1.83 & $-$0.46 & 0.02 & $-$11.73 & N & N \\
2048166$+$342724 & IRAS~20462$+$3416 & 20 48 16.64 & $+$34 27 24.4 & 4.38 & 0.29 & $-$0.12 & $\cdots$ & N & N \\
2050135$+$594551 & IRAS~20490$+$5934 & 20 50 13.58 & $+$59 45 51.2 & 1.61 & 0.63 & $-$0.58 & $\cdots$ & N & N \\
2053379$+$445807 & MSX6C~G085.3935$+$00.1268 & 20 53 37.98 & $+$44 58 07.4 & 0.91 & $-$0.65 & $-$5.98 & $-$26.82 & $\cdots$ & N \\
2057130$+$482200 & IRAS~20555$+$4810 & 20 57 13.08 & $+$48 22 00.3 & 0.72 & 1.72 & $-$0.54 & $-$15.92 & $\cdots$ & N \\
2100253$+$523017 & V\*~V2495~Cyg & 21 00 25.34 & $+$52 30 17.6 & 1.34 & 1.75 & $\cdots$ & $\cdots$ & $\cdots$ & N \\
2101550$+$495135 & IRAS~21002$+$4939 & 21 01 55.02 & $+$49 51 35.5 & 0.77 & 0.97 & $-$0.58 & $-$7.37 & $\cdots$ & N \\
2117391$+$685509 & IRAS~21169$+$6842 & 21 17 39.18 & $+$68 55 09.5 & 1.13 & 0.60 & $-$0.62 & $\cdots$ & $\cdots$ & N \\
2122090$+$492624 & $\cdots$ & 21 22 09.06 & $+$49 26 24.5 & 0.59 & 0.38 & $-$0.81 & $-$9.28 & $\cdots$ & N \\
IRAS~21509$+$6234 & $\cdots$ & 21 52 19.37 & $+$62 48 39.5 & $-$0.41 & $\cdots$ & 0.07 & $\cdots$ & N & N \\
IRAS~21522$+$6018 & $\cdots$ & 21 53 46.10 & $+$60 32 14.2 & $-$0.55 & $\cdots$ & $-$0.14 & $\cdots$ & N & Y \\
2155455$+$575106 & IRAS~21541$+$5736 & 21 55 45.55 & $+$57 51 06.6 & 1.31 & 1.96 & $-$0.79 & $-$8.98 & $\cdots$ & N \\
IRAS~21563$+$5630 & $\cdots$ & 21 58 01.30 & $+$56 44 49.6 & $-$0.58 & $-$1.90 & $-$0.32 & $-$4.45 & N & N \\
2158358$+$585722 & 2MASS~J21583590$+$5857227 & 21 58 35.81 & $+$58 57 22.8 & 0.79 & 1.59 & $-$1.51 & $-$9.08 & $\cdots$ & N \\
2204124$+$530401 & IRAS~22023$+$5249 & 22 04 12.45 & $+$53 04 02.0 & 3.53 & $-$0.17 & $-$0.22 & $-$14.38 & $\cdots$ & N \\
IRAS~22097$+$5647 & $\cdots$ & 22 11 31.88 & $+$57 02 17.4 & $-$0.12 & $-$2.72 & $-$0.15 & $-$4.69 & Y & Y \\
2219055$+$613616 & IRAS~22174$+$6121 & 22 19 05.52 & $+$61 36 16.1 & 0.63 & 1.34 & $-$0.70 & $-$10.34 & $\cdots$ & N \\
2219520$+$633532 & IRAS~22182$+$6320 & 22 19 52.05 & $+$63 35 32.4 & 0.68 & 1.18 & $-$0.74 & $\cdots$ & $\cdots$ & N \\
2233550$+$653918 & $\cdots$ & 22 33 55.02 & $+$65 39 18.5 & 1.12 & 1.73 & $-$6.28 & $\cdots$ & $\cdots$ & N \\
IRAS~22394$+$6930 & $\cdots$ & 22 40 59.80 & $+$69 46 14.7 & $-$0.63 & $\cdots$ & $-$0.08 & $\cdots$ & Y & Y \\
IRAS~22394$+$5623 & $\cdots$ & 22 41 27.10 & $+$56 39 08.0 & $-$0.71 & $\cdots$ & 0.02 & $-$2.58 & $\cdots$ & N \\
2251389$+$515042 & IRAS~22495$+$5134 & 22 51 38.97 & $+$51 50 42.7 & 3.60 & $-$0.22 & $-$0.24 & $\cdots$ & $\cdots$ & N \\
2259442$+$585956 & IRAS~22576$+$5843 & 22 59 44.21 & $+$58 59 56.2 & 1.36 & 1.94 & $-$1.62 & $\cdots$ & $\cdots$ & N \\
2303421$+$614741 & $\cdots$ & 23 03 42.15 & $+$61 47 41.4 & 0.86 & 1.37 & $-$1.39 & $-$17.34 & $\cdots$ & Y \\
2312291$+$612534 & IRAS~23103$+$6109 & 23 12 29.16 & $+$61 25 34.1 & 1.69 & 1.53 & 0.57 & $-$20.58 & $\cdots$ & N \\
2317522$+$580511 & IRAS~23156$+$5748 & 23 17 52.22 & $+$58 05 11.2 & 1.43 & 1.16 & $-$1.05 & $-$9.27 & $\cdots$ & N \\
2335128$+$610005 & IRAS~23328$+$6043 & 23 35 12.84 & $+$61 00 05.4 & 0.97 & 1.80 & $-$2.13 & $-$9.73 & $\cdots$ & N \\
2341559$+$641512 & IRAS~23395$+$6358 & 23 41 55.98 & $+$64 15 12.5 & 0.76 & 1.63 & $-$0.34 & $-$7.54 & $\cdots$ & N \\
2346058$+$632312 & IRAS~23436$+$6306 & 23 46 05.81 & $+$63 23 12.8 & 0.95 & 1.19 & $-$0.27 & $-$10.85 & $\cdots$ & N \\
IRAS~23489$+$6235 & $\cdots$ & 23 51 27.28 & $+$62 51 47.1 & $-$1.44 & $\cdots$ & $-$0.43 & $-$1.84 & $\cdots$ & N \\
IRAS~23554$+$5612 & $\cdots$ & 23 58 01.32 & $+$56 29 13.4 & $-$0.57 & $-$2.19 & $-$0.41 & $\cdots$ & $\cdots$ & N \\
IRAS~23561$+$6037 & $\cdots$ & 23 58 38.70 & $+$60 53 48.0 & $-$0.78 & $\cdots$ & 0.06 & $-$3.13 & $\cdots$ & N \\
\enddata

\tablenotetext{a}{J2000.0.}
\tablenotetext{b}{AC$_{12}$ and AC$_{23}$ represent the AKARI [09]$-$[18] and
                  [18]$-$[65] colours, respectively.
                  See Section~\ref{ssec:objs} for the definition of colour.}
\tablenotetext{c}{$Q1$ and $Q2$ parameters are defined in \citet{messineo12aa},
                  see Section~\ref{ssec:q1q2}.}
\tablenotetext{d}{Detection of OH/H$_{2}$O maser emission in the current 
                  observation is
                  indicated by ``Y''; non-detections are indicated by ``N''. 
                  A blank entry means the object is not observed.}

\end{deluxetable}


\begin{deluxetable}{lcrrrrrrrrcl}
\rotate
\tablewidth{0pt}
\tabletypesize{\scriptsize}
\tablecaption
{Parameters of the OH maser detections.\label{tab:oh}}
\tablehead{
   \colhead{Object} &
   \colhead{Rest Freq.} &
   \colhead{$V_{\rm b,p}$\tablenotemark{a}} &
   \colhead{$F_{\rm b,p}$\tablenotemark{a}} &
   \colhead{$V_{\rm r,p}$\tablenotemark{b}} &
   \colhead{$F_{\rm r,p}$\tablenotemark{b}} &
   \colhead{$V_{\rm b,e}$\tablenotemark{c}} &
   \colhead{$V_{\rm r,e}$\tablenotemark{c}} &
   \colhead{$I$\tablenotemark{d}} &
   \colhead{rms} &
   \colhead{Ref.\tablenotemark{e}} &
   \colhead{Absorptions\tablenotemark{f}} \\

   &
   \colhead{(MHz)} &
   \colhead{(\kms)} &
   \colhead{(Jy)} &
   \colhead{(\kms)} &
   \colhead{(Jy)} &
   \colhead{(\kms)} &
   \colhead{(\kms)} &
   \colhead{(Jy~\kms)} &
   \colhead{(Jy)} &
   &
   \colhead{(\kms)}
}
\startdata
1807272$-$194639 & 1612 & $-$48.4 & 0.62 & $\cdots$ & $\cdots$ & $-$77.4 & 30.4 & 2.71 & 0.03 & new & $\cdots$ \\
                 & 1665 & $\cdots$ & $\cdots$ & $\cdots$ & $\cdots$ & $\cdots$ & $\cdots$ & $\cdots$ & 0.03 & $\cdots$ & 32.2 \\
                 & 1667 & $-$76.3 & 0.88 & $-$45.5 & 0.28 & $-$77.9 & $-$43.3 & 2.69 & 0.03 & 1 & 26.9 \\
IRAS~18056$-$1514 & 1612 & 46.9 & 2.44 & 71.9 & 1.60 & 45.2 & 73.0 & 6.84 & 0.02 & 2 & $\cdots$ \\
                 & 1665 & $\cdots$ & $\cdots$ & $\cdots$ & $\cdots$ & $\cdots$ & $\cdots$ & $\cdots$ & 0.02 & $\cdots$ & $\cdots$ \\
                 & 1667 & $\cdots$ & $\cdots$ & $\cdots$ & $\cdots$ & $\cdots$ & $\cdots$ & $\cdots$ & 0.01 & $\cdots$ & $\cdots$ \\
IRAS~18117$-$1625 & 1612 & 65.1 & 0.31 & 101.4 & 0.19 & 63.9 & 102.5 & 0.55 & 0.03 & new & $\cdots$ \\
                 & 1665 & $\cdots$ & $\cdots$ & $\cdots$ & $\cdots$ & $\cdots$ & $\cdots$ & $\cdots$ & 0.02 & $\cdots$ & $\cdots$ \\
                 & 1667 & $\cdots$ & $\cdots$ & $\cdots$ & $\cdots$ & $\cdots$ & $\cdots$ & $\cdots$ & 0.03 & $\cdots$ & $\cdots$ \\
IRAS~18118$-$1615 & 1612 & 162.7 & 0.24 & $\cdots$ & $\cdots$ & 161.5 & 163.8 & 0.24 & 0.02 & new & $\cdots$ \\
                 & 1665 & $\cdots$ & $\cdots$ & $\cdots$ & $\cdots$ & $\cdots$ & $\cdots$ & $\cdots$ & 0.02 & $\cdots$ & $\cdots$ \\
                 & 1667 & 170.1 & 0.09 & $\cdots$ & $\cdots$ & 160.0 & 178.0 & 0.42 & 0.02 & new & $\cdots$ \\
IRAS~18127$-$1516 & 1612 & $-$18.9 & 0.22 & 18.0 & 0.06 & $-$20.1 & 20.0 & 0.71 & 0.01 & new & $\cdots$ \\
                 & 1665 & $\cdots$ & $\cdots$ & $\cdots$ & $\cdots$ & $\cdots$ & $\cdots$ & $\cdots$ & 0.01 & $\cdots$ & $\cdots$ \\
                 & 1667 & $\cdots$ & $\cdots$ & $\cdots$ & $\cdots$ & $\cdots$ & $\cdots$ & $\cdots$ & 0.01 & $\cdots$ & $\cdots$ \\
1824037$+$063625 & 1612 & $\cdots$ & $\cdots$ & $\cdots$ & $\cdots$ & $\cdots$ & $\cdots$ & $\cdots$ & 0.01 & $\cdots$ & $\cdots$ \\
                 & 1665 & $\cdots$ & $\cdots$ & $\cdots$ & $\cdots$ & $\cdots$ & $\cdots$ & $\cdots$ & 0.01 & $\cdots$ & $\cdots$ \\
                 & 1667 & $-$29.1 & 0.11 & $-$10.4 & 0.12 & $-$30.1 & $-$9.9 & 0.28 & 0.02 & new & $\cdots$ \\
IRAS~18251$-$1048 & 1612 & 71.3 & 1.49 & 111.0 & 2.17 & 67.9 & 112.2 & 10.55 & 0.03 & 3 & 81.5 \\
                 & 1665 & $\cdots$ & $\cdots$ & $\cdots$ & $\cdots$ & $\cdots$ & $\cdots$ & $\cdots$ & 0.04 & $\cdots$ & $\cdots$ \\
                 & 1667 & 23.0 & 0.14 & 113.0 & 0.17 & 9.3 & 126.2 & 3.87 & 0.03 & 3 & 7.1 \\
IRAS~18286$-$0959 & 1612 & $-$0.2 & 0.15 & 12.3 & 0.10 & $-$2.5 & 14.0 & 0.50 & 0.02 & 4 & $\cdots$ \\
                 & 1665 & $\cdots$ & $\cdots$ & $\cdots$ & $\cdots$ & $\cdots$ & $\cdots$ & $\cdots$ & 0.02 & $\cdots$ & 4.2 \\
                 & 1667 & $\cdots$ & $\cdots$ & $\cdots$ & $\cdots$ & $\cdots$ & $\cdots$ & $\cdots$ & 0.02 & $\cdots$ & 6.0 \\
OH~16.3$-$3.0 & 1612 & 20.8 & 8.09 & 35.0 & 6.98 & 11.2 & 45.2 & 41.37 & 0.02 & 4 & $\cdots$ \\
             & 1665 & 21.2 & 0.06 & $\cdots$ & $\cdots$ & 20.1 & 30.0 & 0.31 & 0.01 & new & $\cdots$ \\
             & 1667 & 45.5 & 0.22 & $\cdots$ & $\cdots$ & 24.7 & 46.6 & 1.05 & 0.01 & new & $\cdots$ \\
1838595$-$052024 & 1612 & 41.2 & 0.13 & $\cdots$ & $\cdots$ & 40.1 & 44.1 & 0.23 & 0.02 & new & 67.3, 102.5 \\
                 & 1665 & $\cdots$ & $\cdots$ & $\cdots$ & $\cdots$ & $\cdots$ & $\cdots$ & $\cdots$ & 0.02 & $\cdots$ & $\cdots$ \\
                 & 1667 & 21.4 & 0.07 & $\cdots$ & $\cdots$ & 17.6 & 59.8 & 0.75 & 0.02 & new & $\cdots$ \\
1854158$+$011501 & 1612 & 19.1 & 0.41 & 45.8 & 0.86 & 18.0 & 46.9 & 2.85 & 0.02 & new & $\cdots$ \\
                 & 1665 & 36.6 & 0.07 & $\cdots$ & $\cdots$ & 33.9 & 40.5 & 0.33 & 0.02 & new & $\cdots$ \\
                 & 1667 & 37.3 & 0.13 & $\cdots$ & $\cdots$ & 32.9 & 41.1 & 0.69 & 0.02 & new & 12.6, 29.1, 55.4 \\
1854250$+$004958 & 1612 & 12.3 & 0.14 & $\cdots$ & $\cdots$ & 8.9 & 13.4 & 0.33 & 0.02 & new & $\cdots$ \\
                 & 1665 & $\cdots$ & $\cdots$ & $\cdots$ & $\cdots$ & $\cdots$ & $\cdots$ & $\cdots$ & 0.02 & $\cdots$ & 21.8 \\
                 & 1667 & $\cdots$ & $\cdots$ & $\cdots$ & $\cdots$ & $\cdots$ & $\cdots$ & $\cdots$ & 0.02 & $\cdots$ & 13.1 \\
IRAS~18587$+$0521 & 1612 & $-$15.5 & 0.08 & 33.9 & 0.05 & $-$17.2 & 35.6 & $-$0.58 & 0.01 & new & $\cdots$ \\
                 & 1665 & $\cdots$ & $\cdots$ & $\cdots$ & $\cdots$ & $\cdots$ & $\cdots$ & $\cdots$ & 0.01 & $\cdots$ & $-$14.5, 14.6 \\
                 & 1667 & 24.1 & 0.04 & $\cdots$ & $\cdots$ & 20.8 & 25.8 & 0.16 & 0.01 & new & $\cdots$ \\
1904448$+$042318 & 1612 & 15.7 & 0.13 & 76.4 & 0.32 & 12.3 & 88.9 & 2.27 & 0.02 & new & 20.8 \\
                 & 1665 & 63.5 & 0.20 & $\cdots$ & $\cdots$ & 16.8 & 91.0 & 1.51 & 0.02 & new & 20.1 \\
                 & 1667 & 17.0 & 0.41 & $\cdots$ & $\cdots$ & 12.6 & 91.1 & 2.83 & 0.02 & new & $\cdots$ \\
IRAS~19027$+$0517 & 1612 & 13.4 & 0.15 & $\cdots$ & $\cdots$ & 11.2 & 18.0 & 0.44 & 0.02 & new & 82.1 \\
                 & 1665 & 32.8 & 0.05 & 77.3 & 0.10 & 28.9 & 85.0 & 1.32 & 0.02 & new & $\cdots$ \\
                 & 1667 & 78.4 & 0.14 & $\cdots$ & $\cdots$ & 12.6 & 86.7 & 1.86 & 0.02 & new & $\cdots$ \\
IRAS~19074$+$0534 & 1612 & 12.3 & 0.06 & $\cdots$ & $\cdots$ & 11.1 & 13.4 & 0.07 & 0.01 & new & $\cdots$ \\
                 & 1665 & $\cdots$ & $\cdots$ & $\cdots$ & $\cdots$ & $\cdots$ & $\cdots$ & $\cdots$ & 0.01 & $\cdots$ & 78.3 \\
                 & 1667 & $\cdots$ & $\cdots$ & $\cdots$ & $\cdots$ & $\cdots$ & $\cdots$ & $\cdots$ & 0.01 & $\cdots$ & 76.3 \\
1914408$+$114449 & 1612 & $\cdots$ & $\cdots$ & $\cdots$ & $\cdots$ & $\cdots$ & $\cdots$ & $\cdots$ & 0.02 & $\cdots$ & 21.4, 59.4 \\
                 & 1665 & 59.7 & 0.10 & $\cdots$ & $\cdots$ & 58.0 & 64.1 & 0.29 & 0.02 & new & $\cdots$ \\
                 & 1667 & $\cdots$ & $\cdots$ & $\cdots$ & $\cdots$ & $\cdots$ & $\cdots$ & $\cdots$ & 0.02 & $\cdots$ & $\cdots$ \\
1922557$+$202854 & 1612 & $-$5.3 & 3.68 & 12.9 & 3.15 & $-$8.7 & 14.6 & 14.88 & 0.01 & new & $\cdots$ \\
                 & 1665 & $-$6.8 & 0.07 & 11.9 & 0.12 & $-$7.8 & 13.5 & 0.62 & 0.01 & new & $\cdots$ \\
                 & 1667 & $\cdots$ & $\cdots$ & $\cdots$ & $\cdots$ & $\cdots$ & $\cdots$ & $\cdots$ & 0.01 & $\cdots$ & 7.1 \\
1930003$+$175601 & 1612 & 23.6 & 0.13 & $\cdots$ & $\cdots$ & 20.8 & 25.3 & 0.33 & 0.02 & new & $\cdots$ \\
                 & 1665 & $\cdots$ & $\cdots$ & $\cdots$ & $\cdots$ & $\cdots$ & $\cdots$ & $\cdots$ & 0.02 & $\cdots$ & 25.1 \\
                 & 1667 & $\cdots$ & $\cdots$ & $\cdots$ & $\cdots$ & $\cdots$ & $\cdots$ & $\cdots$ & 0.02 & $\cdots$ & 25.7 \\
IRAS~19312$+$1950 & 1612 & 28.7 & 0.81 & $\cdots$ & $\cdots$ & 10.0 & 36.7 & 3.66 & 0.02 & 5 & $\cdots$ \\
                 & 1665 & 32.8 & 0.10 & $\cdots$ & $\cdots$ & 29.5 & 42.7 & 0.71 & 0.02 & 5 & $\cdots$ \\
                 & 1667 & 30.2 & 0.13 & $\cdots$ & $\cdots$ & 28.0 & 39.0 & 0.81 & 0.02 & 5 & $\cdots$ \\
IRAS~19356$+$0754 & 1612 & $-$119.9 & 3.83 & $\cdots$ & $\cdots$ & $-$138.1 & $-$71.1 & 48.87 & 0.02 & new & $\cdots$ \\
                 & 1665 & $-$94.7 & 0.87 & $\cdots$ & $\cdots$ & $-$134.8 & $-$74.3 & 11.67 & 0.02 & new & $\cdots$ \\
                 & 1667 & $-$133.3 & 0.35 & $\cdots$ & $\cdots$ & $-$138.8 & $-$87.2 & 8.58 & 0.02 & new & $\cdots$ \\
2015573$+$470534 & 1612 & $-$9.3 & 5.15 & $-$0.8 & 2.61 & $-$13.2 & 7.2 & 24.05 & 0.01 & new & $\cdots$ \\
                 & 1665 & $-$6.8 & 0.22 & $-$2.9 & 0.13 & $-$8.4 & $-$1.3 & 0.70 & 0.01 & new & $\cdots$ \\
                 & 1667 & $-$10.4 & 0.98 & $\cdots$ & $\cdots$ & $-$12.6 & 10.4 & 4.46 & 0.01 & new & $\cdots$ \\
IRAS~22097$+$5647 & 1612 & $\cdots$ & $\cdots$ & $\cdots$ & $\cdots$ & $\cdots$ & $\cdots$ & $\cdots$ & 0.01 & $\cdots$ & $\cdots$ \\
                 & 1665 & 63.9 & 0.18 & $-$33.7 & 0.10 & $-$69.4 & $-$28.6 & 0.97 & 0.01 & new & $\cdots$ \\
                 & 1667 & $-$65.3 & 0.64 & $-$43.9 & 0.14 & $-$66.4 & $-$34.0 & 1.87 & 0.01 & new & $\cdots$ \\
IRAS~22394$+$6930 & 1612 & $-$49.6 & 0.13 & $\cdots$ & $\cdots$ & $-$50.7 & $-$47.3 & 0.19 & 0.01 & new & $\cdots$ \\
                 & 1665 & $\cdots$ & $\cdots$ & $\cdots$ & $\cdots$ & $\cdots$ & $\cdots$ & $\cdots$ & 0.01 & $\cdots$ & $\cdots$ \\
                 & 1667 & $\cdots$ & $\cdots$ & $\cdots$ & $\cdots$ & $\cdots$ & $\cdots$ & $\cdots$ & 0.01 & $\cdots$ & $\cdots$ \\
\enddata

\tablenotetext{a}{\vlsr\ and flux density of the blueshifted peak of a
                  double-peaked profile. For a single-peaked or irregular 
                  profile, the brightest peak is recorded in these two 
                  columns, no matter whether it is really ``blueshifted'' 
                  or not.}
\tablenotetext{b}{Same as the above footnote, but for the redshifted peak of a
                  double-peaked profile, if it exists.}
\tablenotetext{c}{\vlsr\ of the two ends of the whole emission profile.
                  The cut-off is defined by the 3-$\sigma$ flux level.}
\tablenotetext{d}{Integrated flux of the whole emission profile.}
\tablenotetext{e}{References for known detections.}
\tablenotetext{f}{\vlsr\ of the absorption features, if any.}
{\bf References.}
(1)~\citet{david93aas}, (2)~\citet{hekkert91aas}, (3)~\citet{engels07aa}, 
(4)~\citet{sevenster01aa}, (5)~\citet{nakashima11apj}
\end{deluxetable}

\begin{deluxetable}{lrrrrrrrrc}
\tablewidth{0pt}
\tabletypesize{\scriptsize}
\tablecaption
{Parameters of the H$_{2}$O maser detections.\label{tab:h2o}}
\tablehead{
   \colhead{Object} &
   \colhead{$V_{\rm b,p}$\tablenotemark{a}} &
   \colhead{$F_{\rm b,p}$\tablenotemark{a}} &
   \colhead{$V_{\rm r,p}$\tablenotemark{b}} &
   \colhead{$F_{\rm r,p}$\tablenotemark{b}} &
   \colhead{$V_{\rm b,e}$\tablenotemark{c}} &
   \colhead{$V_{\rm r,e}$\tablenotemark{c}} &
   \colhead{$I$\tablenotemark{d}} &
   \colhead{rms} &
   \colhead{Ref.\tablenotemark{e}} \\

   &
   \colhead{(\kms)} &
   \colhead{(Jy)} &
   \colhead{(\kms)} &
   \colhead{(Jy)} &
   \colhead{(\kms)} &
   \colhead{(\kms)} &
   \colhead{(Jy~\kms)} &
   \colhead{(Jy)}     
}
\startdata
IRAS~17055$-$0216 & $-$46.7 & 0.49 & $\cdots$ & $\cdots$ & $-$47.5 & $-$45.9 & 0.47 & 0.04 & new \\ 
IRAS~17132$-$0744 & 3.9 & 2.16 & $\cdots$ & $\cdots$ & $-$2.3 & 12.1 & 5.93 & 0.07 & 1 \\ 
IRAS~17171$-$0843 & $-$26.7 & 0.20 & $-$9.1 & 0.48 & $-$27.2 & $-$7.6 & 1.12 & 0.06 & 1 \\ 
IRAS~17343$+$1052 & $-$57.0 & 0.26 & $\cdots$ & $\cdots$ & $-$57.4 & $-$56.6 & 0.16 & 0.05 & 2 \\ 
IRAS~18056$-$1514 & 52.9 & 24.26 & $\cdots$ & $\cdots$ & 50.8 & 63.8 & 62.49 & 0.05 & 1 \\ 
IRAS~18099$-$1449 & $-$1.6 & 0.19 & 8.6 & 0.16 & $-$2.9 & 11.1 & 0.76 & 0.04 & new \\ 
IRAS~18100$-$1250 & 79.8 & 1.25 & $\cdots$ & $\cdots$ & 78.2 & 80.7 & 1.34 & 0.06 & new \\ 
IRAS~18117$-$1625 & 68.5 & 0.15 & $\cdots$ & $\cdots$ & 67.7 & 69.1 & 0.16 & 0.04 & new \\ 
IRAS~18127$-$1516 & $-$15.4 & 4.53 & 13.2 & 0.20 & $-$17.7 & 14.0 & 9.57 & 0.06 & new \\ 
IRAS~18286$-$0959 & 62.3 & 22.41 & $\cdots$ & $\cdots$ & $-$118.9 & 231.0 & 741.01 & 0.05 & 1 \\ 
OH~16.3$-$3.0 & 19.8 & 0.60 & 39.7 & 0.91 & 14.6 & 42.0 & 4.40 & 0.06 & 1 \\ 
IRAS~18455$+$0448 & 21.4 & 0.69 & 46.7 & 2.20 & 14.0 & 50.2 & 4.26 & 0.07 & 1 \\ 
IRAS~18587$+$0521B & $-$26.3 & 0.53 & 18.9 & 0.40 & $-$27.4 & 19.5 & 0.94 & 0.04 & new \\ 
IRAS~19134$+$2131 & $-$41.1 & 0.42 & $-$12.3 & 2.27 & $-$42.0 & $-$8.4 & 5.63 & 0.06 & 3 \\ 
1919572$+$104808 & 31.9 & 0.95 & 53.3 & 0.22 & 31.1 & 53.9 & 1.19 & 0.07 & 4 \\ 
IRAS~19312$+$1950 & 17.3 & 2.44 & $\cdots$ & $\cdots$ & 14.4 & 23.9 & 5.57 & 0.06 & 5 \\ 
IRAS~19356$+$0754 & $-$35.3 & 0.19 & $\cdots$ & $\cdots$ & $-$144.7 & $-$26.2 & 0.45 & 0.01 & new \\ 
IRAS~19464$+$3514 & 33.3 & 0.69 & $\cdots$ & $\cdots$ & 32.3 & 33.9 & 0.62 & 0.09 & 1 \\ 
1951461$+$272458 & 32.3 & 0.48 & $\cdots$ & $\cdots$ & 29.4 & 32.7 & 0.54 & 0.07 & new \\ 
IRAS~20021$+$2156 & 20.2 & 0.22 & $\cdots$ & $\cdots$ & 19.1 & 21.0 & 0.27 & 0.04 & new \\ 
IRAS~21522$+$6018 & 20.4 & 0.32 & $\cdots$ & $\cdots$ & 19.8 & 21.0 & 0.23 & 0.04 & new \\ 
IRAS~22097$+$5647 & $-$49.2 & 54.99 & $\cdots$ & $\cdots$ & $-$60.9 & $-$39.3 & 81.84 & 0.08 & 1 \\ 
IRAS~22394$+$6930 & $-$44.6 & 1.05 & $\cdots$ & $\cdots$ & $-$45.5 & $-$43.2 & 1.18 & 0.05 & new \\ 
2303421$+$614741 & $-$8.4 & 0.51 & 2.3 & 0.19 & $-$9.1 & 2.9 & 0.73 & 0.04 & new \\
\enddata

\tablenotetext{a}{\vlsr\ and flux density of the blueshifted peak of a
                  double-peaked profile. For a single-peaked or irregular 
                  profile, the brightest peak is recorded in these two 
                  columns, no matter whether it is really ``blueshifted'' 
                  or not.}
\tablenotetext{b}{Same as the above footnote, but for the redshifted peak of a
                  double-peaked profile, if it exists.}
\tablenotetext{c}{\vlsr\ of the two ends of the whole emission profile.
                  The cut-off is defined by the 3-$\sigma$ flux level.}
\tablenotetext{d}{Integrated flux of the whole emission profile.}
\tablenotetext{e}{References for known detections.}
{\bf References.}
(1)~Paper~I, (2)~\citet{benson96apjs}, (3)~\citet{imai07apj},
(4)~\citet{engels96aas}, (5)~\citet{nakashima11apj}.
\end{deluxetable}

\begin{deluxetable}{lrrr}
\tablewidth{0pt}
\tablecaption
{Parameters of the OH maser non-detections. The channel spacing of the 
spectrometer was between 0.5\kms\ and 0.6\kms.\label{tab:noh}}
\tablehead{

   \colhead{Object} &
   \multicolumn{3}{c}{rms (Jy)} \\ 
   \cline{2-4}

   & 
   \colhead{1612~MHz} &
   \colhead{1665~MHz} &
   \colhead{1667~MHz}

}
\startdata
IRAS~16030$-$0634 & 0.03 & 0.05 & 0.03 \\
IRAS~16131$-$0216 & 0.02 & 0.02 & 0.02 \\
IRAS~17055$-$0216 & 0.02 & 0.02 & 0.02 \\
IRAS~17132$-$0744 & 0.25 & 0.02 & 0.02 \\
IRAS~17193$-$0601 & 0.03 & 0.02 & 0.02 \\
1750356$-$203743 & 0.02 & 0.02 & 0.02 \\
IRAS~18099$-$1449 & 0.02 & 0.02 & 0.02 \\
IRAS~18100$-$1250 & 0.01 & 0.01 & 0.01 \\
IRAS~19085$+$1038 & 0.02 & 0.02 & 0.02 \\
1911358$+$133111 & 0.02 & 0.01 & 0.01 \\
1912477$+$033435 & 0.02 & 0.02 & 0.02 \\
IRAS~19134$+$2131 & 0.01 & 0.01 & 0.01 \\
1918205$+$014659 & 0.02 & 0.02 & 0.02 \\
1938574$+$103016 & 0.02 & 0.02 & 0.02 \\
1951461$+$272458 & 0.02 & 0.02 & 0.02 \\
IRAS~20010$+$2508 & 0.01 & 0.01 & 0.01 \\
2003357$+$284847 & 0.02 & 0.02 & 0.02 \\
2003599$+$351617 & 0.02 & 0.02 & 0.02 \\
IRAS~20021$+$2156 & 0.01 & 0.01 & 0.01 \\
2008383$+$410040 & 0.01 & 0.01 & 0.01 \\
IRAS~20215$+$6243 & 0.01 & 0.01 & 0.01 \\
IRAS~20305$+$6246 & 0.02 & 0.02 & 0.02 \\
2040444$+$465322 & 0.02 & 0.02 & 0.02 \\
2048044$+$390459 & 0.02 & 0.02 & 0.02 \\
2048166$+$342724 & 0.02 & 0.02 & 0.01 \\
2050135$+$594551 & 0.02 & 0.02 & 0.02 \\
IRAS~21509$+$6234 & 0.01 & 0.01 & 0.01 \\
IRAS~21522$+$6018 & 0.02 & 0.02 & 0.02 \\
IRAS~21563$+$5630 & 0.01 & 0.01 & 0.01 \\
\enddata
\end{deluxetable}


\begin{deluxetable}{lr}
\tablewidth{0pt}
\tablecaption
{Parameters of the H$_{2}$O maser non-detections. The channel spacing of the 
spectrometer was about 0.8\kms.\label{tab:nh2o}}
\tablehead{

   \colhead{Object} &
   \colhead{rms (Jy)}

}
\startdata
0038592$+$592746 & 0.06 \\
0122182$+$634055 & 0.06 \\
0128404$+$632737 & 0.06 \\
0253373$+$691539 & 0.06 \\
0358076$+$624425 & 0.06 \\
0413220$+$501428 & 0.07 \\
0433014$+$343840 & 0.09 \\
0536468$+$314600 & 0.08 \\
0540502$+$340241 & 0.08 \\
0547450$+$003842 & 0.09 \\
0608452$+$130841 & 0.08 \\
IRAS~16030$-$0634 & 0.07 \\
IRAS~16131$-$0216 & 0.06 \\
IRAS~17193$-$0601 & 0.06 \\
1741385$-$241435 & 0.10 \\
1750356$-$203743 & 0.09 \\
1752536$-$184100 & 0.08 \\
1807272$-$194639 & 0.03 \\
IRAS~18118$-$1615 & 0.06 \\
1824037$+$063625 & 0.07 \\
1824288$-$155108 & 0.08 \\
IRAS~18362$-$0521 & 0.05 \\
1838595$-$052024 & 0.07 \\
1848010$+$000448 & 0.07 \\
1854158$+$011501 & 0.07 \\
1854250$+$004958 & 0.04 \\
IRAS~18587$+$0521A & 0.04 \\
IRAS~18596$+$0605 & 0.04 \\
1904448$+$042318 & 0.07 \\
IRAS~19027$+$0517 & 0.04 \\
IRAS~19074$+$0534 & 0.05 \\
IRAS~19085$+$1038 & 0.04 \\
1911358$+$133111 & 0.07 \\
1912477$+$033435 & 0.09 \\
1914408$+$114449 & 0.07 \\
1918205$+$014659 & 0.10 \\
1922250$+$131851 & 0.10 \\
1922557$+$202854 & 0.04 \\
1923002$+$151051 & 0.11 \\
1930003$+$175601 & 0.04 \\
1938574$+$103016 & 0.09 \\
IRAS~20010$+$2508 & 0.04 \\
2003357$+$284847 & 0.09 \\
2003599$+$351617 & 0.10 \\
2008383$+$410040 & 0.09 \\
2009217$+$271859 & 0.10 \\
2013142$+$370536 & 0.09 \\
2015573$+$470534 & 0.04 \\
2020151$+$364334 & 0.04 \\
IRAS~20215$+$6243 & 0.04 \\
2029222$+$403543 & 0.04 \\
IRAS~20305$+$6246 & 0.04 \\
2033464$+$450840 & 0.08 \\
2040444$+$465322 & 0.08 \\
2048044$+$390459 & 0.09 \\
2048166$+$342724 & 0.05 \\
2050135$+$594551 & 0.05 \\
2053379$+$445807 & 0.08 \\
2057130$+$482200 & 0.08 \\
2100253$+$523017 & 0.08 \\
2101550$+$495135 & 0.04 \\
2117391$+$685509 & 0.07 \\
2122090$+$492624 & 0.04 \\
IRAS~21509$+$6234 & 0.04 \\
2155455$+$575106 & 0.04 \\
IRAS~21563$+$5630 & 0.05 \\
2158358$+$585722 & 0.05 \\
2204124$+$530401 & 0.05 \\
2219055$+$613616 & 0.05 \\
2219520$+$633532 & 0.05 \\
2233550$+$653918 & 0.10 \\
IRAS~22394$+$5623 & 0.05 \\
2251389$+$515042 & 0.05 \\
2259442$+$585956 & 0.05 \\
2312291$+$612534 & 0.09 \\
2317522$+$580511 & 0.06 \\
2335128$+$610005 & 0.06 \\
2341559$+$641512 & 0.07 \\
2346058$+$632312 & 0.08 \\
IRAS~23489$+$6235 & 0.09 \\
IRAS~23554$+$5612 & 0.10 \\
IRAS~23561$+$6037 & 0.09 \\
\enddata
\end{deluxetable}

\end{document}